# Turbulent buoyant confined jet with variable source temperature

M. F. El-Amin[1,2], Amgad Salama[1] and Shuyu Sun[1]

[1] King Abdullah University of Science and Technology (KAUST), Thuwal 23955-6900, Kingdom of Saudi Arabia
[2] Department of Mathematics, Faculty of Science, Aswan University, Aswan 81528, Egypt

**Abstract**

In this work, experimental and numerical investigations are considered for confined buoyant turbulent jet with varying inlet temperatures. Results of the experimental work and numerical simulations for the problem under consideration are presented. Four cases of different variable inlet temperatures and different flow rates are considered. The realizable $k-\varepsilon$ turbulence model is used to model the turbulent flow. Comparisons show good agreements between simulated and measured results. The average deviation of the simulated temperature by realizable $k-\varepsilon$ turbulent model and the measured temperature is within 2%. The results indicate that temperatures along the vertical axis vary, generally, in nonlinear fashion as opposed to the approximately linear variation that was observed for the constant inlet temperature that was done in a previous work. Furthermore, thermal stratification exits particularly closer to the entrance region. Further away from the entrance region the variation in temperatures becomes relatively smaller. The stratification is observed since the start of the experiment and continues during whole time. Numerical experiments for constant, monotone increasing and monotone decreasing of inlet temperature are done to show its effect on the buoyancy force in terms of Richardson number.

# Nomenclature

$A_0$    constant defined in Eq. (23)

$A_s$    constant defined in Eq. (24)

$b$    local jet width [m]

CFD    **C**omputational **F**luid **D**ynamics

$c$    lateral spread rate of the jet

$C_1$    parameter defined in Eq. (20)

$C_2$    constant equals 1.9

$C_p$    specific heat [m²/s²K]

$C_{1\varepsilon}$    constant equals 1.44

$C_{3\varepsilon}$    parameter defined in Eq. (20)

$C_\mu$    coefficient defined in Eq. (21)

$d$    inlet diameter [m]

$g$    acceleration due to gravity [m/s²]

$G_b$    generation of turbulence due to buoyancy

$G_k$    generation of turbulent kinetic energy due to the mean velocity gradient

$I_0$    intensity of turbulence at the nozzle inlet

$k$    turbulent kinetic energy

$r$    radial axis [m]

$R$    Richardson number

Re    Reynolds number

$S$    strain rate

$T$    temperature [K]

| | |
|---|---|
| $T_{exp}$ | measured temperature [K] |
| $T_{in}$ | inlet temperature [K] |
| $T_{sim}$ | simulated temperature [K] |
| $t$ | time [s] |
| $p$ | pressure [Pa] |
| $P$ | dynamic pressure [Pa] |
| $P_0$ | inlet dynamic pressure [Pa] |
| $P_c$ | centerline dynamic pressure [Pa] |
| $Pr_t$ | turbulent Prandtl number for energy |
| $u$ | mean axial velocity component [m/s] |
| $u_c$ | centerline velocity [m/s] |
| $u_0$ | inlet velocity [m/s] |
| $v$ | mean radial velocity component [m/s] |
| $z$ | axial axis [m] |

**Greek symbols**

| | |
|---|---|
| $\Delta t$ | time step [s] |
| $\Delta T$ | temperature difference [K] |
| $\Delta \rho$ | density difference [kg/m$^3$] |
| $\beta$ | thermal expansion coefficient [1/K] |
| $\varepsilon$ | turbulence dissipation rate |
| $\lambda$ | thermal conductivity W/(m.K) |
| $\mu$ | dynamic viscosity [kg/m s] |
| $\mu_t$ | turbulent viscosity |

| $\nu$ | kinematics viscosity [m² s] |
|---|---|
| $\rho$ | density [kg/m³] |
| $\theta$ | constant defined in Eq. (25) |
| $\sigma_k$ | turbulent Prandtl number for $k$ |
| $\sigma_\varepsilon$ | turbulent Prandtl number for $\varepsilon$ |

**Introduction**

Although stratification of fluids due to the existence of temperature gradients is not desirable in many processes that require homogenization, it is, in other processes (e.g., heat storage tanks) desirable because of the low mixing mechanisms involved which help maintaining the required temperature distribution. The problem, however, in thermally stratified heat storage tanks is its sensitivity to external disturbance. That is the state of thermal stratification could be destroyed once sufficient disturbance is introduced. In particular, the conditions at the inlet are considered as one example of such disturbance sources. Therefore, it is important to study the effect of inlet conditions on stratification behavior of such systems. Since, in heat storage tanks and in many other applications, fluids enter the tank in the form of buoyant jet, great deal of works on jet flow have been considered either experimentally or numerically. However, the problem of buoyant heated jet with variable source temperature which can be found in many industrial and environmental applications, has received relatively little attention. Jet flow can be divided mainly into three types, pure jet, pure plume and forced plume. In pure jets, fluids inter the domain with high momentum fluxes which essentially cause higher intensity of turbulent mixing. In pure plume, on the other hand, buoyancy fluxes cause local acceleration leading to turbulent mixing. In the general case of a forced plume a combination of initial momentum and buoyancy fluxes are responsible for turbulent mixing. Several techniques were developed to study jets in confined spaces as will be explained later. Lately, with the increase of computers efficiencies and capacities, computational fluid dynamics (CFD) became one of the essential tools to exploring on fluids behavior of such fundamental importance. In turbulent flows Reynolds Average Navier-Stokes technique (RANS) are usually adopted in order to make the system amenable to solution. The problem of using RANS approach, however, is that till now, there is no unifying set of equation to model all kinds of turbulent flows and heat transfer scenarios. Therefore, it is important to choose the model which suites the case under investigation and even to calibrate its coefficients in order to fit experimental results. El-Amin et al. [1] investigated the 2D upward, axisymmetric turbulent confined jet and developed several models to describe flow patterns using realizable $k-\varepsilon$ turbulence model. Furthermore, CFD analysis of the flow structure of a horizontal water jet entering a rectangular tank has been done by El-Amin et al. [2, 3]. Their findings were later used by Panthalookaran et al. [4] to calibrate both realizable and RNG $k-\varepsilon$ turbulence models so that they may be used for simulating stratified hot water storage tanks.

Comprehensive reviews of jet flows were presented by Rajaratnam [5] and List [6]. Furthermore several experimental works were conducted to highlight the interesting patterns and the governing parameters pertinent to this kind of flows including the work of Wygnanski and Fiedler [7], Rodi [8], Panchapakesan and Lumley [9-10], Fukushima et al. [11] Agrawal et al. [12], O'Hern et al. [19] and many others. On the other hand, several phenomena pertinent to buoyant jets were investigated by many authors. For example, the problem of entrainment by a plume or jet at density interface was considered by Baines [13], mechanisms involved in transition to turbulence in buoyant plume flow was investigated by Kmura and Bejan [14], round buoyant jets were also investigated by Shabbir and George [15], and Papanicolaou and List [16], bifurcation in a buoyant horizontal laminar jet was studied by Arakeri, et al. [17]. With respect to the kind of fluids used in buoyant jet studies, several researchers have considered different fluids for either experimental or numerical investigations. For example O'Hern et al. [18] performed experimental work on a turbulent buoyant helium plume. El-Amin and Kanayama [19, 20] studied buoyant jet resulting from hydrogen leakage.

They developed the similarity formulation and solutions of the centerline quantities such as velocity and concentration. Furthermore, El-Amin [21] introduced a numerical investigation of a vertical axisymmetric non-Boussinesq buoyant jet resulting from hydrogen leakage in air as an example of injecting a low-density gas into high-density ambient. On the other hand, the mechanics of buoyant jet flows issuing with a general three-dimensional geometry into an unbounded ambient environment with uniform density or stable density stratification and under stagnant or steady sheared current conditions is investigated by Jirka [22]. He formulated an integral model for the conservation of mass, momentum, buoyancy and scalar quantities in the turbulent jet flow. Furthermore Jirka [23] extended this work to also encounter plane buoyant jet dynamics resulting from the interaction of multiple buoyant jet effluxes spaced along a diffuser line. In the previous work by El-Amin et al. [1], analyses of the components of 2D axisymmetric vertical unheated/heated turbulent confined jet using turbulence realizable $k - \varepsilon$ model were conducted. Moreover experimental work was elaborated for temperature measurements of such system to provide verification of the models used. In that work, several models were considered to describe axial velocity, centerline velocity, radial velocity, dynamic pressure, mass flux, momentum flux and buoyancy flux for both unheated (non-buoyant) and heated (buoyant) jet. In that work inlet temperatures were considered fixed. However, in many applications inlet temperatures are not, generally fixed. An experimental study of a stratified thermal storage under variable inlet temperature for different inlet designs was performed by Abo-Hamdan et al. [25]. Furthermore, Yoo and Kim [24] introduced approximate analytical solutions for stratified thermal storage under variable inlet temperature.

In this work, analysis of vertical hot water jet entering a cylindrical tank filled with cold water with variable inlet temperature is conducted. The inlet temperature of the buoyant jet is allowed to change within a small range and is considered as a function of time. Numerical investigations under the above mentioned conditions are performed in order to obtain fields of pressure, velocity, temperature and turbulence. 2D axisymmetric simplification is assumed to reduce the grid size in the solution domain and the realizable $k-\varepsilon$ model is used to model turbulent flow.

**Measurements**

Schematic diagram of the experimental setup is shown in Fig. 1. The cylindrical tank, made of 0.005 m thick galvanized iron sheets, has a diameter of 0.36 m and a height of 0.605 m is shown in Fig. 2a. The inlet pipe is located at center of the bottom of the tank with an inner diameter of 0.02 m. the inlet pipe is inserted in the tank up to a height of 0.06 m above the base of the tank. The outlet pipe, located at center of the top of the tank, has an inner diameter of 0.02 m with a depth in the tank of 0.055 m from the top plate. The uncertainty in the diameter of the inlet and outlet pipes is $\pm$ 0.001 m. This geometry suggests that the tank, the inlet, and the outlet may be modeled as axisymmetric around the vertical axis. Thermal effects are measured by thermocouples of K-type which were calibrated against a standard PT-25 resistance thermometer with an average calibration error of $\pm 0.25$ K. Flow rate is measured using a magnetic-type with a calibration error $\pm 3.5\%$. The temperature is recorded in Kelvin using a data acquisition system connected with a personal computer. The data acquisition system has an error about $\pm 1$ K. The above estimated errors are included in the measured data. Thermal effects are measured by inserting a vertical rod with 9 stainless steel sheathed K-type thermocouples. The nine sensors are distributed at different heights as, 0.06, 0.12, 0.18, 0.24, 0.30, 0.36, 0.42, 0.48 and 0.54 m measured from the bottom. The distance between the symmetry axis and the thermocouples' rod is 0.09 m, i.e. in the middle between the symmetry axis and the tank wall. The inlet temperature was measured using another thermocouple in which is located at the inlet pipe. It is important to indicate that all precautions have been taken to make sure the geometrical symmetry is achieved in the sense of the alignment of the inlet and outlet tubes, the smoothness of tube materials, the inlet geometry, etc.

For the variable inlet temperature, the used parameters are listed in Table 1. The duration for measurements for each case was approximately 30 min. The inlet temperature and the initial temperature are given in column 2 and 3 of the Table 1, respectively, and the flow rate is given in column 4. The inlet velocity, Reynolds number and turbulence intensity at the inlet nozzle are calculated from the given data.

The best fitting for the given curves of the inlet variable temperature, Fig. 3a, can be represented by a 5$^{th}$ order polynomial as a function of time, Eqs. (A.1-A.4) in the Appendix. The corresponding Reynolds numbers, Fig. 3b, and inlet turbulence intensity, Fig. 3c, are calculated as functions of inlet temperature which in turn have been represented by 5$^{th}$ order polynomials of time, Eqs. (A.5-A.8) and (A.9-A.12), respectively, with the aid of Eqs. (13)-(14). These functions may be represented by other polynomials with less order but the deviation from the measured data will increase.

The following empirical relation (Fluent User's Guide, Fluent Inc. 2003, ch. 6) is used to describe the turbulent intensity at the inlet nozzle as a function of the Reynolds number:

$$I_0 = 0.16(\text{Re})^{-0.125} \qquad (13)$$

The Reynolds number with the inlet diameter as a length scale is defined by the relation:

$$\text{Re} = \frac{u_0 d}{\nu} \qquad (14)$$

The measured temperature profiles for variable inlet temperature are plotted as a function of time, at different sensors positions (cases 1-4) in Fig. 4 (a-d). It is apparent that the temperatures along the vertical axis vary in nonlinear fashion with time, especially, at larger heights $z \geq 0.18$ m (plume region). Also, the figures indicate that the temperature increases as time increases. Thermal stratification is observed looking at the difference in temperature top (higher) to bottom (lower). The degree of stratification, however, seems to be more pronounced in the lower half of the tank than in the top half. The stratification is verified from the beginning of the experiment and continues during whole time.

**Mathematical Formulation**

A comparison study was done by El-Amin et al. [1] to test different turbulence models when simulating confined buoyant jet, and they reported that the best model is the realizable $k - \varepsilon$ model. Therefore, in this work we consider this model to simulate the problem under consideration. The realizable $k - \varepsilon$ model developed by Shih et al. [26] involves a new eddy-viscosity formula originally proposed by Reynolds [27] and a new model equation for dissipation $\varepsilon$ based on the dynamic equation of the mean-square vorticity fluctuations. The Reynolds-averaged Navier-Stokes equations (RANS) are given in Eqs. (2) - (3), and the energy equation is represented by Eq. (4). The governing equations of mass, momentum and turbulence take the form:

Continuity equation:

$$\frac{\partial u}{\partial z} + \frac{1}{r}\frac{\partial (rv)}{\partial r} = 0 \tag{1}$$

Axial momentum equation:

$$\frac{\partial u}{\partial t} + u\frac{\partial u}{\partial z} + \frac{v}{r}\frac{\partial (ru)}{\partial r} = -\frac{1}{\rho}\frac{\partial p}{\partial z} + \frac{1}{\rho}\frac{\partial}{\partial z}\left[(\mu + \mu_t)\left(2\frac{\partial u}{\partial z}\right)\right] + \\ \frac{1}{\rho r}\frac{\partial}{\partial r}\left[r(\mu + \mu_t)\left(\frac{\partial u}{\partial r} + \frac{\partial v}{\partial z}\right)\right] + \beta g(T - T_0) - \frac{2}{3}\frac{\partial k}{\partial z} \tag{2}$$

Radial momentum equation:

$$\frac{\partial v}{\partial t} + u\frac{\partial v}{\partial z} + \frac{v}{r}\frac{\partial (rv)}{\partial r} = -\frac{1}{\rho}\frac{\partial p}{\partial r} + \frac{1}{\rho}\frac{\partial}{\partial z}\left[(\mu + \mu_t)\left(\frac{\partial u}{\partial r} + \frac{\partial v}{\partial z}\right)\right] + \frac{1}{\rho r}\frac{\partial}{\partial r}\left[r(\mu + \mu_t)\left(2\frac{\partial v}{\partial r}\right)\right] - \frac{2}{3r}\frac{\partial (rk)}{\partial r} \quad (3)$$

Energy equation:

$$\frac{\partial T}{\partial t} + u\frac{\partial T}{\partial z} + \frac{v}{r}\frac{\partial (rT)}{\partial r} = \frac{1}{\rho}\frac{\partial}{\partial z}\left[(\frac{\lambda}{c_p} + \frac{\mu_t}{\Pr_t})\left(\frac{\partial T}{\partial z}\right)\right] + \frac{1}{\rho r}\frac{\partial}{\partial r}\left[r(\frac{\lambda}{c_p} + \frac{\mu_t}{\Pr_t})\left(\frac{\partial T}{\partial r}\right)\right] \quad (4)$$

Turbulent kinetic energy ($k$) equation:

$$\frac{\partial k}{\partial t} + u\frac{\partial k}{\partial z} + \frac{v}{r}\frac{\partial (rk)}{\partial r} = \frac{1}{\rho}\frac{\partial}{\partial z}\left[(\mu + \frac{\mu_t}{\sigma_k})\frac{\partial k}{\partial z}\right] + \frac{1}{\rho r}\frac{\partial}{\partial r}\left[r(\mu + \frac{\mu_t}{\sigma_k})\frac{\partial k}{\partial r}\right] + \frac{1}{\rho}\left[G_k + G_b - \rho\varepsilon\right] \quad (5)$$

Turbulence dissipation rate ($\varepsilon$) equation:

$$\frac{\partial \varepsilon}{\partial t} + u\frac{\partial \varepsilon}{\partial z} + \frac{v}{r}\frac{\partial (r\varepsilon)}{\partial r} = \frac{1}{\rho}\frac{\partial}{\partial z}\left[(\mu + \frac{\mu_t}{\sigma_\varepsilon})\frac{\partial \varepsilon}{\partial z}\right] + \frac{1}{\rho r}\frac{\partial}{\partial r}\left[r(\mu + \frac{\mu_t}{\sigma_\varepsilon})\frac{\partial \varepsilon}{\partial r}\right] + C_1 S\varepsilon - C_2\frac{\varepsilon^2}{k + \sqrt{\nu\varepsilon}} + C_{1\varepsilon}C_{3\varepsilon}G_b\frac{\varepsilon}{k} \quad (6)$$

In the above equations, $u$ and $v$ are the mean axial and radial velocity components, respectively. The other quantities are time, $t$, density $\rho$, acceleration due to gravity, $g$, pressure, $p$, dynamic viscosity, $\mu$, kinematics viscosity, $\nu$, thermal conductivity, $\lambda$, turbulent Prandtl number for $k$, $\sigma_k$, turbulent Prandtl number for $\varepsilon$, $\sigma_\varepsilon$, $T$ is the temperature, $T_0$ is the reference operating temperature.

The turbulence dissipation rate is denoted by $\varepsilon$, while $k$ is the turbulent kinetic energy of the turbulent fluctuations per unit mass. The turbulent viscosity $\mu_t$ is defined as:

$$\mu_t = \rho C_\mu k^2 / \varepsilon \quad (7)$$

where $C_\mu$ is coefficient, which is a new variable defined in the realizable $k - \varepsilon$ model and given by the relation:

$$C_\mu = \varepsilon/(A_0 \varepsilon + A_s kS) \tag{8}$$

where

$$A_0 = 4.04, \tag{9}$$

$$A_s = 6^{1/2} \cos\theta, \tag{10}$$

$$\theta = (1/3)\cos^{-1}(6^{1/2} S^2), \tag{11}$$

$$S = 0.5(\partial v/\partial z + \partial u/\partial r) \tag{12}$$

The eddy viscosity formulation is based on the realizability constraints, the positivity of the normal stress and Schwarz inequality for turbulent shear stresses.

Furthermore, in Eq. (20), $C_1$ defined by the form:

$$C_1 = \max[0.43, Sk/(Sk + 5\varepsilon)] \tag{13}$$

and

$$G_k = \mu_t S^2 \tag{14}$$

is the generation of turbulent kinetic energy due to the mean velocity gradient.

$$C_{3\varepsilon} = \tanh|u/v| \tag{15}$$

The velocity component $u$ parallel to the gravitational vector and $v$ is the component of the velocity perpendicular to the gravitational vector. In this way, $C_{3\varepsilon} = 1$ for buoyant shear layers for which the main flow direction is aligned with the direction of gravity (the case under study). For buoyant shear layers that are perpendicular to the gravitational vector, $C_{3\varepsilon} = 0$.

The generation of turbulence due to buoyancy is given by the relation:

$$G_b = \beta g \frac{\mu_t}{\Pr_t} \frac{\partial T}{\partial z} \tag{16}$$

$\text{Pr}_t$ is the turbulent Prandtl number for energy and $\beta$ is the thermal expansion coefficient. The model constants of the $k-\varepsilon$ model are established to ensure that the model performs well for certain canonical flows such as pipe flow, jet flow, and boundary layer flow. $C_{1\varepsilon}=1.44,\ C_2=1.9,\ \sigma_k=1.0,\ \sigma_\varepsilon=1.2$ and $\text{Pr}_t=0.85$.

Shih et al. [26] have compared their model (realizable $k-\varepsilon$ turbulence model) with experimental data as well as with the standard $k-\varepsilon$ model for a round jet flow and other flows. The comparison shows a good matching between their model and the experimental data than the standard model. The realizable $k-\varepsilon$ model implies that the model satisfies specific constraints on the Reynolds's stresses that make the model more consistent with the physics of turbulent flows and hence more accurate than the other turbulent model. This model contains a new transport equation for the turbulent dissipation rate. Also, a critical coefficient of the model, $C_\mu$, is expressed as a function of mean flow and turbulence properties, rather than assumed to be constant as in the standard model. This allows the model to satisfy certain mathematical constraints on the normal stresses consistent with the physics of turbulence (realizability). Additionally, the realizable $k-\varepsilon$ model uses different sources and sinks terms in the transport eddy dissipation. The modified prediction of $\varepsilon$ along with the modified calculation of $\mu_t$, makes this model superior to the other $k-\varepsilon$ models. For the jet flow this model does better in predicting the spreading rate especially, for near region z<0.35 (see El-Amin et al. [1]).

Boundary conditions need to be specified on all surfaces of the computational domain. Boundaries presented in this study include inflow (Inlet), outflow (outlet), solid wall and axis of symmetry as shown in Fig. 2b. $\Omega_{in}, \Omega_{out}$ and $\Omega_{wall}$ denote the boundary of the inlet, outlet and wall, respectively. In addition to the non-realistic boundary on the axis of symmetry $\Omega_{axis}$. The velocity-inlet boundary conditions imposed at the nozzle are,

$$u = u_0, v = 0, \text{ and } T = T_{in} \text{ on } \Omega_{in} \tag{17}$$

$T_{in}$ is defined in Eq. (A.1-A.4) for the cases of variable inlet temperature.

Due to the stagnant conditions of water inside the tank before the beginning of the influx, all velocity components were initially set to zero. Heat transfer through the walls of the tank is not taken into consideration (adiabatic walls). Density of water, specific heat, thermal conductivity and viscosity are formulated as a piecewise-linear profile of temperature. The turbulence intensity and hydraulic diameter characterizes the turbulence at the inlet boundary. The following equation of empirical correlation for pipe flows is used to describe the turbulence intensity at the inlet boundary as a function of the Reynolds number,

$$I_0 = 0.16(\text{Re})^{-0.125} \quad \text{on } \Omega_{in} \tag{18}$$

Alternatively, one can use the following k and ε on the inlet boundary as (see Kadem et al. [28]),

$$k = k_{in} = 0.005 u_0^2, \quad \varepsilon = \varepsilon_{in} = k_{in}^{3/2} / (0.3(d/2)) \text{ on } \Omega_{in} \tag{19}$$

where *d* is nozzle diameter.

The boundary condition on the axis of symmetry is represented by free-slip condition which is a non-realistic wall with no-friction when velocity and other components near the wall are not retarded. Unlike the no-slip boundary condition for which flow has zero velocity in the wall, free-slip flow is tangent to the surface. On the axis of symmetry, the radial velocity component v, and the gradient of the other dependent variables (*u, ε, k, T*) were equal to zero. So, one may write them as,

$$\frac{\partial u}{\partial r} = 0,\ v = 0,\ \frac{\partial T}{\partial r} = 0,\ \frac{\partial k}{\partial r} = 0,\ \frac{\partial \varepsilon}{\partial r} = 0 \text{ on } \Omega_{axis} \tag{20}$$

Solid wall boundary conditions are represented along the solid walls; the no-slip boundary condition for velocities, zero value for turbulent kinetic energy, and zero gradients for temperature and energy dissipation rate were used.

$$u = 0, v = 0,\text{ on } \Omega_{wall},\ \frac{\partial T}{\partial n_1} = 0,\ k = 0,\ \frac{\partial \varepsilon}{\partial n_1} = 0,\text{ on } \Omega_{wall} \tag{21}$$

where $n_1$ is the outward normal of the wall.

Finally, the outlet boundary which water discharged outside it freely, can be formulated as,

$$\frac{\partial u}{\partial n_2} = 0,\ \frac{\partial v}{\partial n_2} = 0,\ \frac{\partial T}{\partial n_2} = 0,\ \frac{\partial k}{\partial n_2} = 0,\ \frac{\partial \varepsilon}{\partial n_2} = 0 \text{ on } \Omega_{out} \tag{22}$$

where $n_2$ is the outward normal of the outlet boundary.

Also, static pressure can be defined at a known given point in the domain and Fluent extrapolates all other conditions from the interior of the domain.

Initial conditions are described as follows,

$$u = 0, v = 0, T = T_0, k = 0, \varepsilon = 0 \quad \text{at} \ t = 0 \tag{23}$$

In fact, very small values are given as initial conditions for $k$ and $\varepsilon$ instead of zero which only speed up convergence of the solution.

## Numerical Investigations

Fig. 2b shows the computational domain with dimensions of: radius=0.18 m and height=0.605 m. The radius of the inlet and outlet pipes is 0.01 m, while the inlet height inside the tank is 0.06 m and the outlet depth in the tank is 0.055 m. The meshes are built up of Quadratic Submap cells. The number of grid elements used for all calculations is 7,984. Fluent 6.1 and the grid generation tool Gambit [29] are used to model the flow in the tank by solving the continuity, momentum, turbulence and energy equations.

In order to prove grid independence, numerical experiment for case 4, is repeated on the systematically refined grids of sizes 7,984 (grid-1), 9,240 (grid-2), 12,880 (grid-3) and 23,560 (grid-4) quadrilateral cells, respectively. The minimum distances between the nodes in the respective grids are 0.00125 m, 0.002 m, 0.0014285 m and 0.001 m and the maximum distances between the nodes are 0.0055 m, 0.004583 m, 0.00366 m and 0.00275 m respectively in the order of refinement. Figs. 5 (a, b) show the results of the grid refinement studies for the axial velocity and Temperature, respectively. The maximum deviation caused by grid is about 3 % for the velocity, and 0.14 % for the temperature which could be negligible.

In order to achieve convergence, Under-Relaxation is applied on pressure, velocities, energy, turbulent viscosity, turbulence kinetic energy and turbulent dissipation rate calculations. Body Force Weighted Discretization is used for pressure and the velocity-pressure coupling is treated using the SIMPLE algorithm. A Second-Order Upwind scheme is used in the equations of momentum, energy, turbulence kinetic energy and turbulence dissipation rate. Segregated Implicit Solver with the Implicit Second-Order scheme is used.

In order to use a suitable time step, we performed a comparison for one case with different time steps as 0.01, 0.1, 0.5 and 1 s which are shown in Table 2. This comparison includes temperature, axial velocity, turbulent kinetic energy and turbulent dissipation rate of kinetic energy. From this table one can note that the differences are negligible values. Then, to reduce the time of calculation we have to use the time step of 1 sec.

## Comparisons

Both measured and simulated temperatures as a function of the tank height for various times and variable inlet temperatures (cases 1-4) are plotted in Figs. 6 (a-d), respectively. Good agreement between the experimental and numerical data is observed. The maximum error observed is 0.35 K, however, for cases 1, 2, 3 and 4 the maximum error is 0.2, 0.2, 0.35 and 0.35 K, respectively, occurs after 30 min of charging process.

## Axial and Radial Velocities

The mean positive axial velocity $u$ (excluding the reflected velocity with the negative values) is normalized by the centerline velocity $u_c$ against $r/cz$ ($r$ normalized by the jet width $b=cz$, $c$ is the lateral spread rate of the jet) with different heights, for the case 4 at 15 min, is plotted in Fig. 7. It can be seen from this figure that axial velocity profile shows self-similar behavior. Therefore, axial velocity may be represented by a Gaussian distribution using centerline velocity, $u_c$, height, $z$, and width, $b$, as parameters. The Gaussian function takes the form:

$$u = u_c \exp\left(-\frac{r^2}{b^2}\right) \tag{24}$$

This empirical model is plotted in Fig. 8 with comparison with the simulated axial velocity. In this study, the parameter of lateral spread rate of the jet $c$=0.11 which lies in the range of the standard values as reported by Fischer et al. [30]. One can note a relatively large error at small velocities at the both ends of the bell-shape curve.

Using axial velocity definition, Eq. (24), centerline axial velocity (velocity at the axis of symmetry) can be given as:

$$u_c = u_0 A_u / (z - z_0) \tag{25a}$$

such that $u_c = u(r \rightarrow 0)$.

Alternatively, the centerline velocity may be written in the form:

$$u_c = u_0 B_u d / (z - z_0), \tag{25b}$$

to be comparable with the common formula of the centerline velocity given in literature. It is notable that $A_u = B_u d$, $B_u$ specifies the decay rate of the time averaged centerline velocity.

Dimensional arguments together with experimental observations suggest that the mean flow variables, which are known as similarity solutions, are conforming with Eqs. (25) (Fisher et al. [30], Hussein et al. [31], and Shabbir and George [32]). The continuity equation, Eq. (1), for the time-averaged velocities can be solved by substituting the axial velocity form into Eq. (1) to obtain the cross-stream radial velocity in the form:

$$\frac{v}{u_c} = \frac{c}{\eta}\left[-\frac{5}{6} + \frac{5}{6}\exp(-\eta^2) + \eta^2 \exp(-\eta^2)\right] \tag{26}$$

where, $\eta = r/b(z) = r/c(z - z_0)$

## Dynamic Pressure

The dynamic pressure behaves similar to the axial velocity but of course it is scalar quantity such that we do not see negative beaks of the curve. The dynamic pressure can be defined according to the equation:

$$P_d = \frac{1}{2}\rho(u^2 + v^2) \tag{27}$$

At inlet $(u,v) = (u_0, 0)$, therefore, $P_0 = \frac{1}{2}\rho u_0^2$ is the jet nozzle dynamic pressure. On the other hand, one can model the simulated dynamic pressure by the relation:

$$P_d = P_c \exp\left(-\frac{r^2}{h^2 z^2}\right) \tag{28a}$$

or

$$P_d = P_c \exp\left(-\frac{2r^2}{b^2}\right) \tag{28b}$$

where $P_c$ is the centerline dynamic pressure, and $h = c/\sqrt{2}$.

Figure 9 illustrates a comparison between the simulated dynamic pressure and its Gaussian fitting using Eq. (28) as a function of $r$ for different positions of $z$ of the unheated jet at $t=5$ min (case 4). This figure shows a good matching for this Gaussian distribution of the dynamic pressure.

## Selected Simulated Results

In Fig. 10 temperature profiles are plotted against $z$ at different times. One may notice relatively higher temperatures closer to the inlet up to, approximately, $z \approx 0.12$ m, and then it decreases as z increases. As the time proceeds, the temperature closer to the inlet decreases as shown in the figure while it increases further away. The turbulence intensity as a function of the axis of symmetry $z$ for various times is shown in Fig. 11. The turbulence intensity is defined as the ratio of the root-mean-square of the turbulent velocity fluctuations and the mean velocity. Apparently closer to the inlet velocity fluctuations increases due to the impingement of the jet in the relatively quiescent fluid in the tank. However, away from the inlet the intensity of turbulence decreases because of the decrease in velocity as the jet spreads laterally as manifested in Fig. 12. For $z \leq 0.14$ m the turbulence intensity is the same during all time duration, while for $z > 0.14$ m the turbulence intensity decreases with time. It is interesting to note that inside the outlet pipe the turbulence intensity increases as manifested by the sharp increase in turbulence intensity towards the outlet pipe as a result of the influence of pipe wall.

The velocity magnitude as a function of radial axis distance, $r$, at different positions of $z$ at $t$=10 min is plotted in Fig. 12. The velocity magnitude in bottom part of the tank is larger closer to the axis of symmetry while it has smaller values far from it (i.e., as $r$ increases). As $z$ increases velocity magnitude decreases closer to axis of symmetry $z$ while it increases as $r$ increases. This behavior may be explained by the fact that the jet leaves the inlet with a higher velocity and disperses laterally as it moves far from the source. Figure 13 shows temperature as a function of $r$ for various values of $z$ at $t$=10 min. At the bottom of the tank (i.e., small z), the temperature is higher closer to the symmetry axis and it is sharply decreases far from it (i.e., as $r$ increases). Therefore, as $z$ increases and the jet disperses more laterally, the temperature closer to the axis of symmetry decreases while increasing as $r$ increases.

## Jet Richardson Number

Richardson number is defined as a ratio of the buoyancy and the inertia forces. But for more convenience we will define the Richardson number according to the local centerline velocity.

Richardson number is calculated using buoyancy-related terms (density difference) and the velocity at the same point. In jet flows, Richardson number takes the form, [30]:

$$R = \left(\frac{\pi}{4}\right)^{1/2}\left(\frac{g\Delta\rho\, d}{u_c^2}\right)^{1/2} \tag{29}$$

Richardson number is plotted in Fig. 14 against the height z, at different times for Case 3. From this figure it can be seen that Richardson number is reduced in the region closer to the nozzle inlet, and then it increases with the height. In the bottom part the inertia effect dominates the buoyancy effect (jet-like zone), therefore Richardson number decreases. In the top part of the tank, on the other hand, the buoyancy effect dominates the inertia (plume-like zone) as manifested by the increase of Richardson number. Also, in this zone Richardson number increases with time because temperature increases with time and enhances the buoyancy while it decreases closer to the inlet as the inlet temperature is set to decrease.

In order to examine the effect of varying the inlet temperature on Richardson number we perform three numerical experiments, one of them with constant inlet temperature, and two with monotony increasing and monotony decreasing inlet temperature, respectively. The inlet temperatures for these numerical experiments are defined as:

$T_{in} = 308.9539 \text{ K},$ for the constant inlet temperature,

$T_{in} = 301.5519 + 0.01 \times t^2,$ for the monotone increasing inlet temperature,

$T_{in} = 308.9539 - 0.01 \times t^2,$ for the monotone decreasing inlet temperature,

where, $1 \leq t \leq 30$ min, for monotone increasing inlet temperature, $301.552 \leq T_{in} \leq 309.962$ and for monotone decreasing inlet temperature, $308.954 \leq T_{in} \leq 300.554$.

Figure 15 shows Richardson number for the case of constant inlet temperature. From this figure it can be seen that Richardson number behavior is similar for all times closer to the inlet (i.e., in the jet-like region). In the plume-like region Richardson number increases with time because of the increase in temperature. Figures 16 and 17 illustrate Richardson number for the monotone increasing and monotone decreasing inlet temperature, respectively. One can notice that Richardson number in the plume-like region in the case of monotone increasing inlet temperature increases with time as a manifestation of the increased buoyancy, Fig. 16. On the other hand, for the monotone decreasing inlet temperature, Fig.17, Richardson number decreases in the Jet-like region as a manifestation of the decreased temperature.

**Conclusions**

This work is devoted to investigate the problem of non-uniform inlet temperature of buoyant jet. An analysis for vertical hot water jets entering a cylindrical tank filled with cold water under the condition of variable inlet temperature is introduced. The variable inlet temperature is considered as a function of time of charging process. Experimental measurements are performed for the different cases in sequential time steps for both constant and variable inlet temperature. Numerical investigations under the above mentioned conditions are performed. The realizable $k - \varepsilon$ turbulence model is used to simulate turbulent flow for this problem. Comparisons between the measured and simulated temperature show good agreements. Selected empirical Gaussian model with standard parameters are used to represent the simulated results. Selected simulated quantities such as velocity magnitude, temperature and turbulence intensity are investigated. The results indicate that temperature varies, approximately, linearly with time for the constant inlet temperature cases, while, it seems to be, approximately, polynomial or logarithms functions of time for the variable inlet temperature, especially, for plume region. Also, thermal stratification exits; however thermal layers in top part of the tank thinner than them in the bottom part. The stratification is verified from the beginning of experiment and continues during whole time.

**Acknowledgement**

The first author would like to thank the Alexander von Humboldt (AvH) Foundation, Germany for funding his fellowship and for supporting of this research project.

**References:**

1. El-Amin MF, Sun S, Heidemann W, Müller-Steinhagen H (2010) Analysis of a turbulent buoyant confined jet modeled using realizable $k-\varepsilon$ model. Heat Mass Transfer, 46(8): 943-960.

2. El-Amin MF, Heidemann W, Müller-Steinhagen H (2005) Turbulent jet flow into a water store. Proc. Heat Transfer in Components and Systems for Sustainable Energy Technologies, 5-7 April 2005, Grenoble, France, 345-349.

3. El-Amin MF, Heidemann W, Müller-Steinhagen H (2004) Unsteady buoyancy-induced and turbulent flow from a hot horizontal jet entrance into a solar water storage. WSEAS Int. Conf. Heat and Mass Transfer (HMT 2004), Corfu Island, Greece, Aug. 17-19, 2004.

4. Panthalookaran V, El-Amin MF, Heidemann W, Müller-Steinhagen H (2008) Calibrated models for simulation of stratified hot water heat stores. Int. J. Energy Res. 32: 661–676.

5. Rajaratnam N (1976) Turbulent jets. Elsevier Science, New York

6. List EJ (1982) Turbulent jets and plumes. Annu Rev Fluid Mech 14:189–212

7. Wygnanski I, Fiedler H (1969) Some measurements in a self-preserving jet. J Fluid Mech 38:577–612

8. Rodi W (1975) A new method of analyzing hot-wire signals in highly turbulent flow and its evaluation in a round jet. DISA Information 17, February 1975

9. Panchapakesan NR, Lumley JL (1993) Turbulence measurements in axisymmetric jets of air and helium. Part 1. Air jet. J Fluid Mech 246:197–223

10. Panchapakesan NR, Lumley JL (1993) Turbulence measurements in axisymmetric jets of air and helium. Part 2. Helium jet. J Fluid Mech 246:225–247

11. Fukushima C, Aanen L, Westerweel J (2000) Investigation of the mixing process in an axisymmetric turbulent jet using PIV and LIF. 10th International symposium on applications of laser techniques to fluid mechanics, 13–16 July, Lisbon, Portugal

12. Agrawal A, Prasad AK (2003) Integral solution for the mean flow profiles of turbulent jets, plumes, and wakes. ASME J Fluids Eng 125:813–822

13. Baines WD (1975) Entrainment by a plume or jet at a density interface. J Fluid Mech 68(2):309–320

14. Kmura S, Bejan A (1983) Mechanism for transition to turbulence in buoyant plume flow. Int J Heat Mass Transf 26:1515–1532

15. Shabbir A, George K (1994) Experiments on a round turbulent buoyant plume. J Fluid Mech 215:1–32

16. Papanicolaou PN, List EJ (1988) Investigations of round vertical turbulent buoyant jets. J Fluid Mech 195:341–391

17. Arakeri JH, Das D, Srinivasan J (2000) Bifurcation in a buoyant horizontal laminar jet. J Fluid Mech 412:61–73

18. O'Hern TJ, Weckman EJ, Gerhart AL, Tieszen SR, Scefer RW (2005) Experimental study of a turbulent buoyant helium plume. J Fluid Mech 544:143–171

19. El-Amin MF, Kanayama H (2009) Integral solutions for selected turbulent quantities of small-scale hydrogen leakage: a non-buoyant jet or momentum-dominated buoyant jet regime. Int J Hydrogen Energy 34:1607–1612

20. El-Amin MF, Kanayama H (2009) Similarity consideration of the buoyant jet resulting from hydrogen leakage. Int J Hydrogen Energy 34:5803–5809

21. El-Amin MF (2009) Non-Boussinesq turbulent buoyant jet resulting from hydrogen leakage in air. Int J Hydrogen Energy 34:7873–7882

22. Jirka GH (2004) Integral model for turbulent buoyant jets in unbounded stratified flows. Part 1: single round jet. Environ Fluid Mech 4:1–56

23. Jirka GH (2006) Integral model for turbulent buoyant jets in unbounded stratified flows. Part 2: plane jet dynamics resulting from multiport diffuser jets. Environ Fluid Mech 6:43–100

24. Yoo H, Kim CJ, Kim CW (1999) Approximate analytical solutions for stratified thermal storage under variable inlet temperature. Solar Energy 66:47-56

25. Abo-Hamdan MG, Zurigat YH, and Ghaja, AJ (1992) An experimental study of a stratified thermal storage under variable inlet temperature for different inlet designs. Int. J. Heat Mass Transfer 35:1927-1934

26. Shih TH, Liou WW, Shabbir A, Yang Z, Zhu J (1995) A new $k-\varepsilon$ eddy-viscosity model for high Reynolds number turbulent flows-model development and validation. Comput Fluids 24(3):227–238

27. Reynolds WC (1987) Fundamentals of turbulence for turbulence modeling and simulation. Lecture notes for Von Karman institute, Agard Report No. 755

28. Kadem K, Mataoui A, Salem A, Younsi R (2007) Numerical simulation of heat transfer in an axisymmetric turbulent jet impinging on a flat plate. Adv Model Optim 9(2):207–217

29. Fluent 6.1 (2003) User's Guide, Fluent Inc.

30. Fischer HB, List EJ, Koh RCY, Imberger J, Brooks NH (1979) Mixing in inland and coastal waters. Academic Press, San Diego

31. Hussein JH, Capp SP, George WK (1994) Velocity measurements in a high-Reynolds-number, momentum-conserving, axisymmetric, turbulent jet. J Fluid Mech 258:31–75

32. Shabbir A, George K (1994) Experiments on a round turbulent buoyant plume. J Fluid Mech 215:1–32

## Appendix:

The inlet variable temperature may be given as a function of time for each case as follows:

$$T_6(t) = 299.79 + 1.3959\,t - 0.3036\,t^2 + 0.0261\,t^3 - 0.0011\,t^4 + 2\times 10^{-5}\,t^5 \tag{A.1}$$

$$T_7(t) = 307.18 + 0.4911\,t - 0.1411\,t^2 + 0.0097\,t^3 - 0.0003\,t^4 + 3\times 10^{-6}\,t^5 \tag{A.2}$$

$$T_8(t) = 302.74 + 2.9018\,t - 0.4661\,t^2 + 0.0292\,t^3 - 0.0008\,t^4 + 9\times 10^{-6}\,t^5 \tag{A.3}$$

$$T_9(t) = 298.09 + 5.1479\,t - 0.7869\,t^2 + 0.0508\,t^3 - 0.0015\,t^4 + 2\times 10^{-5}\,t^5 \tag{A.4}$$

These polynomials are plotted in Fig. 3a. The ranges of variation of the inlet temperature are:

$$302.04 \geq T_6(t) \geq 297.81 \quad K$$

$$308.15 \geq T_7(t) \geq 300.73 \quad K$$

$$308.95 \geq T_8(t) \geq 301.55 \quad K$$

$$309.43 \geq T_9(t) \geq 301.50 \quad K$$

The corresponding Reynolds numbers are:

$$\mathrm{Re}_6(t) = 1182.1 + 38.618\,t - 8.4158\,t^2 + 0.7241\,t^3 - 0.0312\,t^4 + 7\times 10^{-4}\,t^5 \tag{A.5}$$

$$\mathrm{Re}_7(t) = 3037.8 + 37.278\,t - 20.685\,t^2 + 1.9567\,t^3 - 0.0918\,t^4 + 21\times 10^{-4}\,t^5 \tag{A.6}$$

$$\mathrm{Re}_8(t) = 2679.6 + 293.99\,t - 58.476\,t^2 + 5.1016\,t^3 - 0.2298\,t^4 + 52\times 10^{-4}\,t^5 \tag{A.7}$$

$$\mathrm{Re}_9(t) = 2344.7 + 480.08\,t - 92.483\,t^2 + 8.3257\,t^3 - 0.3938\,t^4 + 94\times 10^{-4}\,t^5 \tag{A.8}$$

These polynomials are shown in Fig. 3b. The ranges of variation of the Reynolds numbers are:

$$1244 \geq Re_6(t) \geq 1130$$

$$3144 \geq Re_7(t) \geq 2239$$

$$3201 \geq Re_8(t) \geq 2715$$

$$3235 \geq Re_9(t) \geq 2711$$

Also, the nozzle inlet turbulence intensity can be represented in a polynomial form of time:

$$I_{06}(t) = 0.0661 - 0.0003t + 6 \times 10^{-5} t^2 - 5 \times 10^{-6} t^3 + 2 \times 10^{-7} t^4 - 4 \times 10^{-9} t^5 \tag{A.9}$$

$$I_{07}(t) = 0.0587 - 0.0002t + 5 \times 10^{-5} t^2 - 5 \times 10^{-6} t^3 + 2 \times 10^{-7} t^4 - 5 \times 10^{-9} t^5 \tag{A.10}$$

$$I_{08}(t) = 0.0596 - 0.0007t + 1 \times 10^{-4} t^2 - 1 \times 10^{-5} t^3 + 5 \times 10^{-7} t^4 - 1 \times 10^{-8} t^5 \tag{A.11}$$

$$I_{09}(t) = 0.0605 - 0.0012t + 2 \times 10^{-4} t^2 - 2 \times 10^{-5} t^3 + 1 \times 10^{-6} t^4 - 2 \times 10^{-8} t^5 \tag{A.12}$$

These profiles are illustrated in Fig. 3c. The ranges of variation of the inlet turbulence intensities are:

$$0.0664 \geq I_{06}(t) \geq 0.0657$$

$$0.0597 \geq I_{07}(t) \geq 0.0585$$

$$0.0596 \geq I_{08}(t) \geq 0.0583$$

$$0.0596 \geq I_{09}(t) \geq 0.0583$$

**Table Captions**

**Table 1:** Summary of the experimental data with variable inlet temperature

**Table 2:** Time step comparison of temperature, axial velocity, turbulent kinetic energy and turbulent dissipation rate of kinetic energy, for case 4

# Figure Captions

**Fig. 1:** Schematic diagram of the experimental setup.

**Fig. 2 (a, b):** Schematic representation of the calculation domain.

**Fig. 3 (a, b, c):** Variable inlet (a) temperature, (b) Reynolds number and (c) turbulence intensity as a function of time for cases 1-4.

**Fig. 4 (a, b, c, d):** Profiles of measured temperature as a function of time, at different sensors positions, for variable inlet temperature, cases 1-4.

**Fig. 5 (a, b):** Grid independence test by (a) axial velocity, and (b) temperature.

**Fig. 6 (a, b, c, d):** Comparison between measured and simulated temperature as function of tank height for variable inlet temperature (cases 1-4).

**Fig. 7:** Normalized axial velocity as a function of r/cz at different positions of z of case 4 at t=15 min.

**Fig. 8:** Comparison between simulated and empirical Gaussian model of axial velocity as a function of r for different positions of z of the case 4 at t=15 min.

**Fig. 9:** Comparison between the simulated dynamic pressure and its Gaussian fitting as a function of r for different positions of z of the case 4 at t=15 min.

**Fig. 10:** Temperature as a function of the axis of symmetry z(r=0) with varies times, case 1.

**Fig. 11:** Turbulence intensity as a function of the axis of symmetry z(r=0) with varies times, case 1.

**Fig. 12:** Velocity magnitude as a function of r with varies values of z at t=10 min, case 3.

**Fig. 13:** Temperature as a function of r with varies values of z at t=10 min, case 3.

**Fig. 14:** Richardson number as a function of the height z, at different times, for case 3.

**Fig. 15:** Richardson number as a function of the height z, at different times, with constant inlet temperature.

**Fig. 16:** Richardson number as a function of the height z, at different times, with monotone increasing inlet temperature.

**Fig. 17:** Richardson number as a function of the height z, at different times, with monotone decreasing inlet temperature.

**Table 1:**

| Case | Inlet temperature [K] | Initial temperature [K] | Flow rate [m³/s] | Inlet velocity [m/s] | Re [-] | Turbulence intensity [%] |
|---|---|---|---|---|---|---|
| 1 | $T_6(t)$ | 294.11 | 0.0000161 | 0.051 | $Re_6(t)$ | $I_{06}(t)$ |
| 2 | $T_7(t)$ | 292.18 | 0.0000161 | 0.051 | $Re_7(t)$ | $I_{07}(t)$ |
| 3 | $T_8(t)$ | 293.77 | 0.0000139 | 0.044 | $Re_8(t)$ | $I_{08}(t)$ |
| 4 | $T_9(t)$ | 292.98 | 0.0000111 | 0.035 | $Re_9(t)$ | $I_{09}(t)$ |

**Table 2:**

| $\Delta t$ | 0.01 s | 0.1 s | 0.5 s | 1 s |
|---|---|---|---|---|
| $T$ | 322.815 | 322.815 | 322.815 | 322.815 |
| $u$ | 0.081836 | 0.081836 | 0.081835 | 0.081835 |
| $k$ | 0.0000244 | 0.0000244 | 0.0000244 | 0.0000244 |
| $\varepsilon$ | 1.33E-05 | 1.33E-05 | 1.33E-05 | 1.33E-05 |

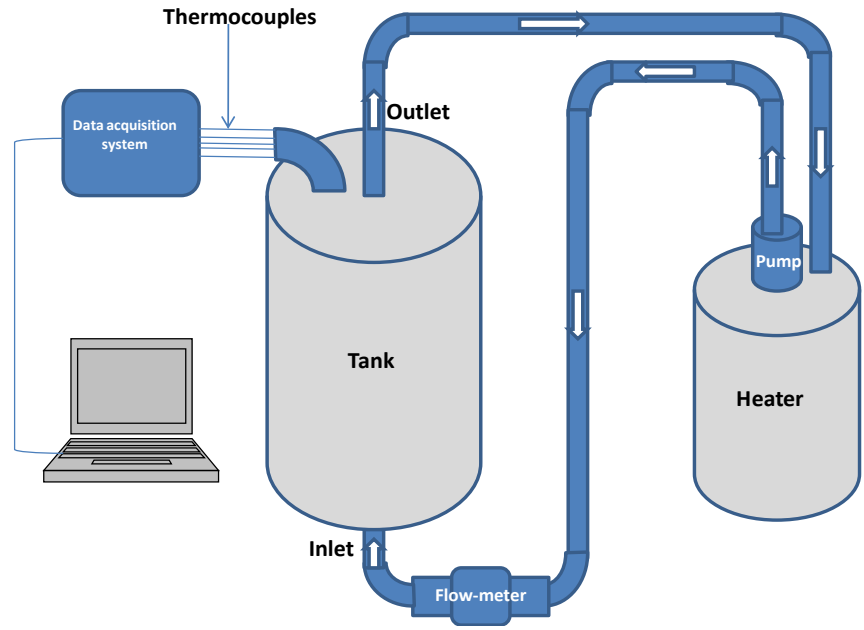

**Fig. 1**

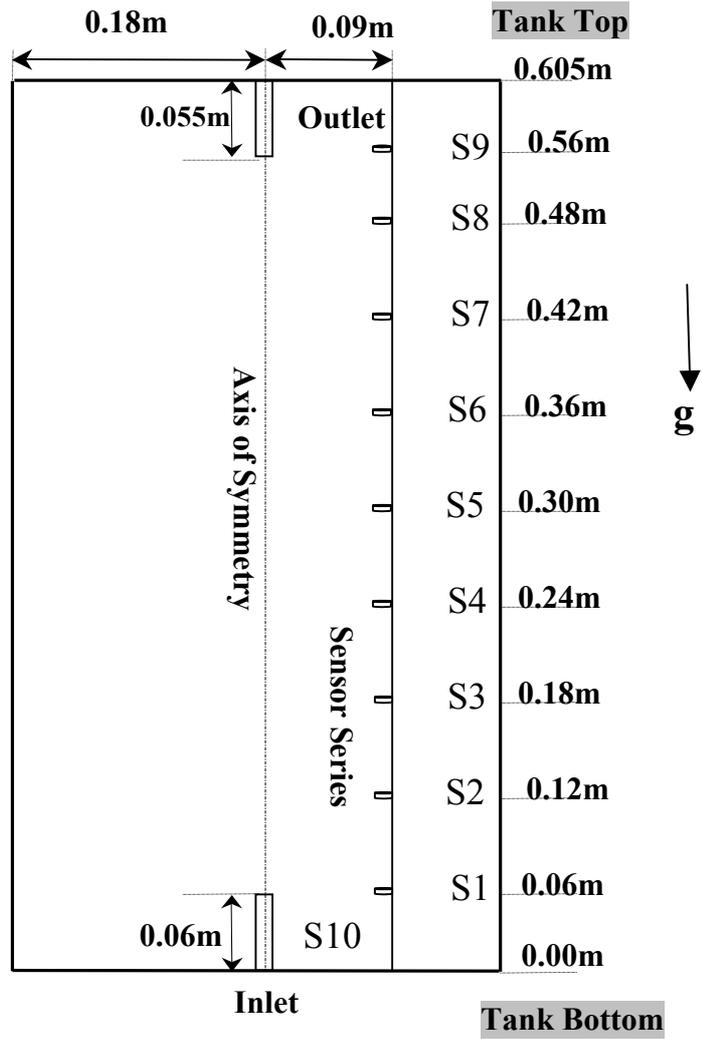

**Fig. 2a**

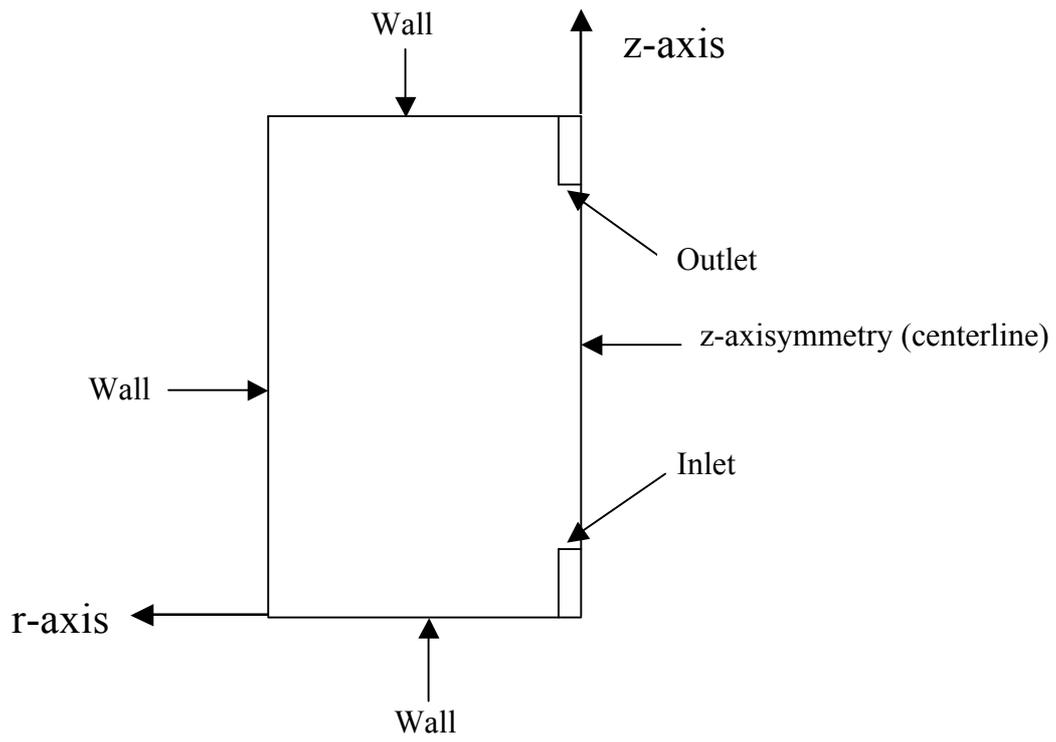

**Fig. 2b**

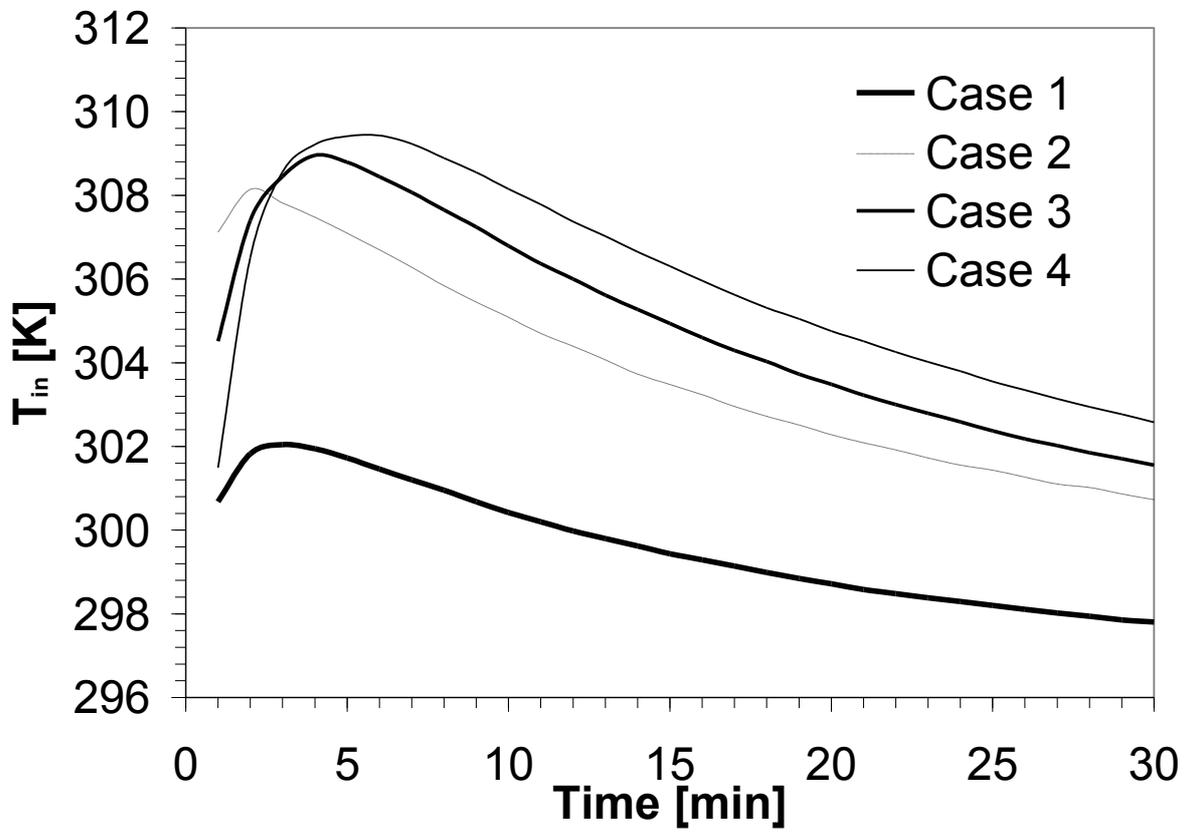

**Fig. 3a**

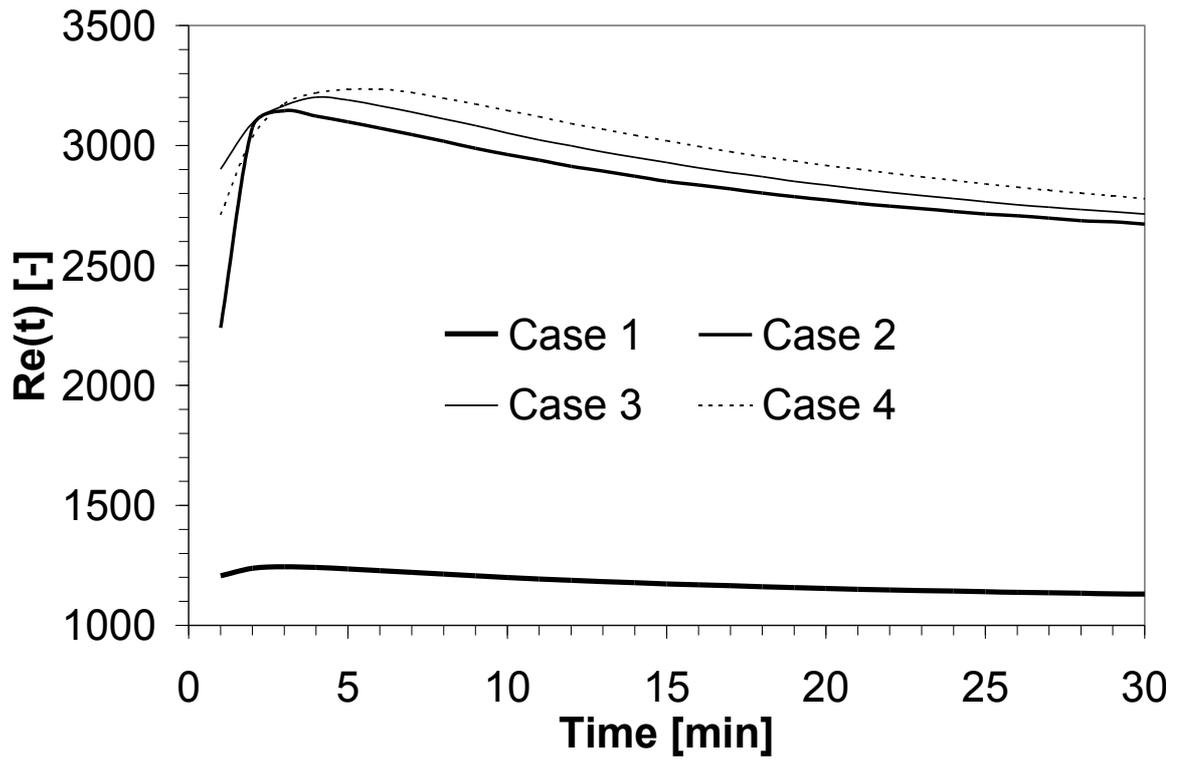

**Fig. 3b**

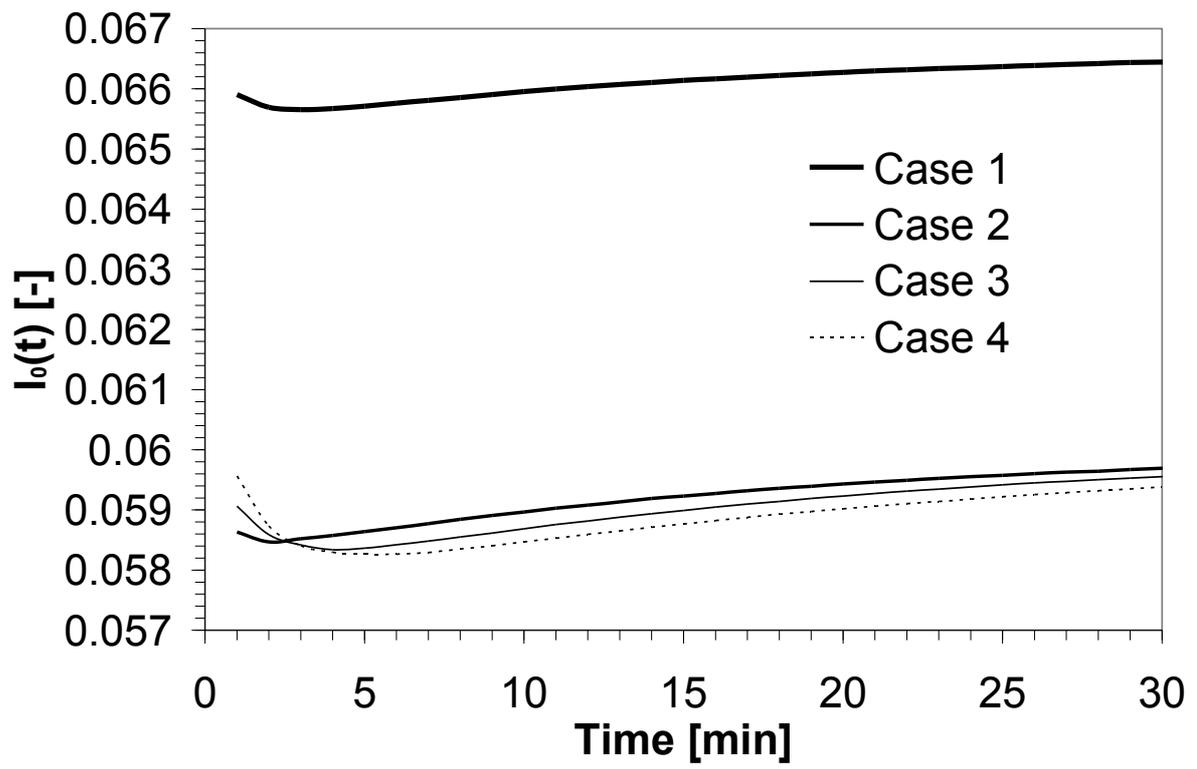

**Fig. 3c**

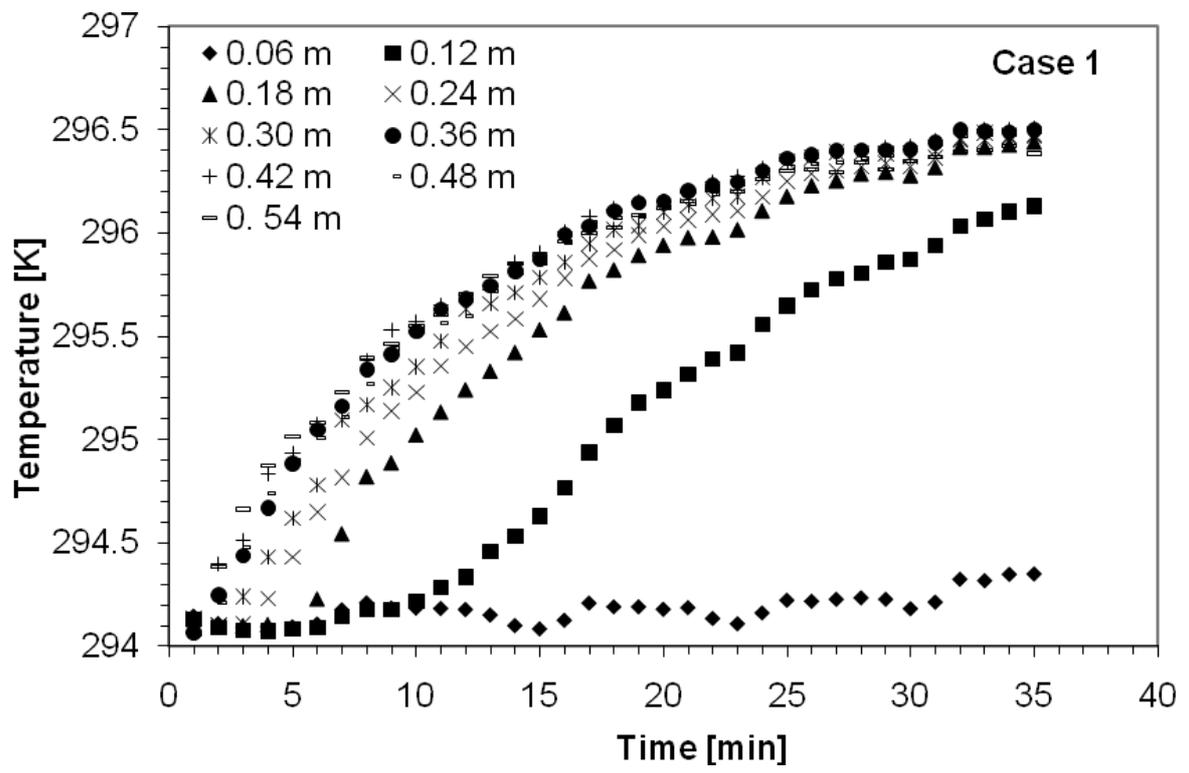

**Fig. 4a**

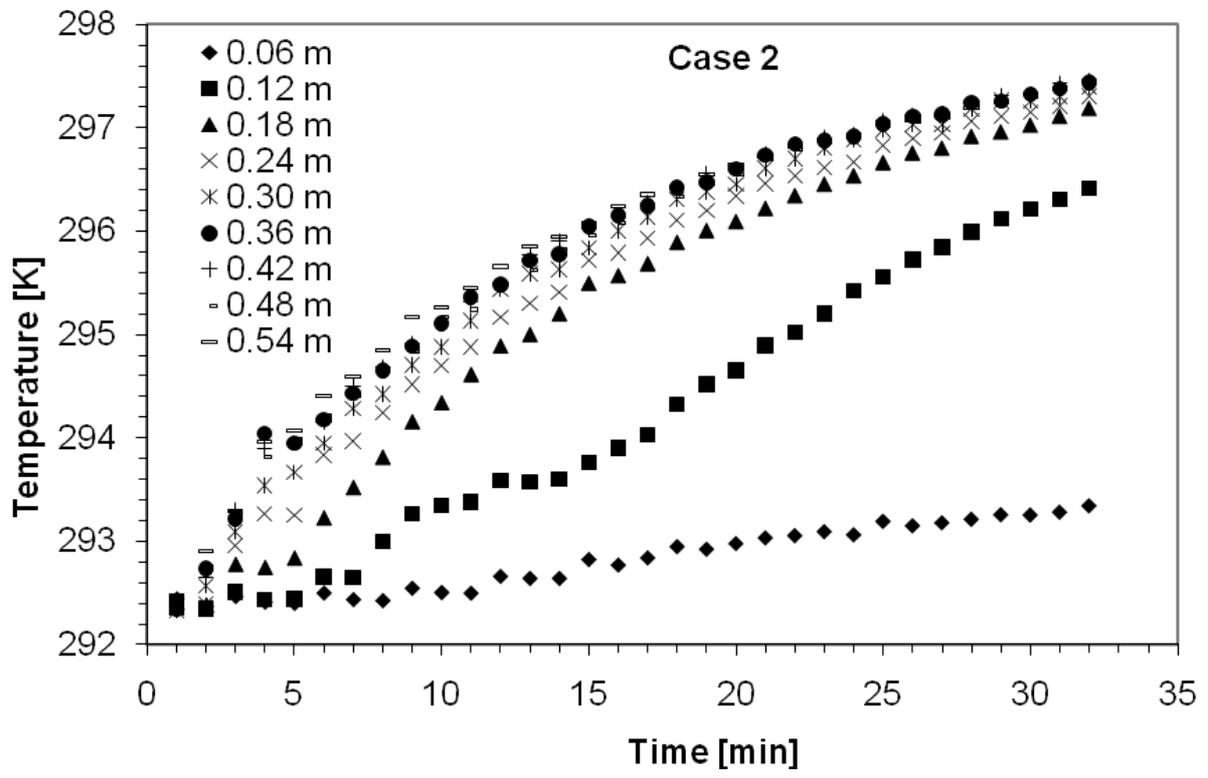

**Fig. 4b**

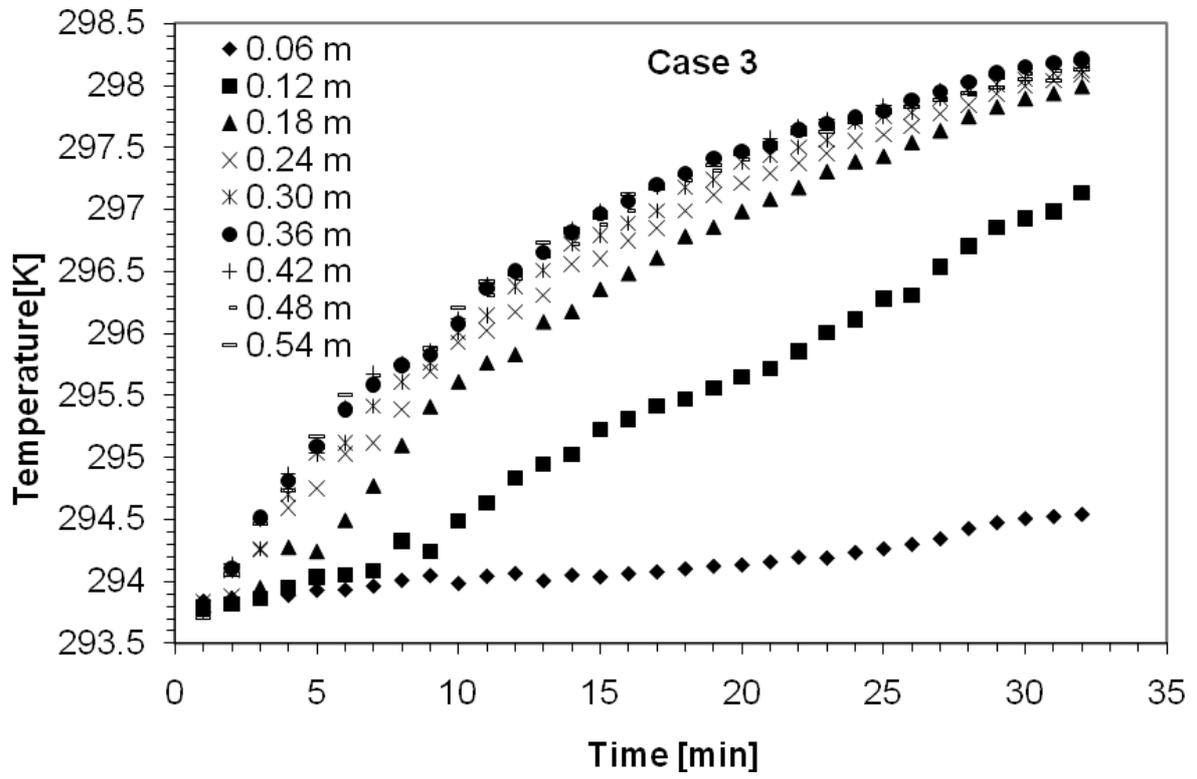

**Fig. 4c**

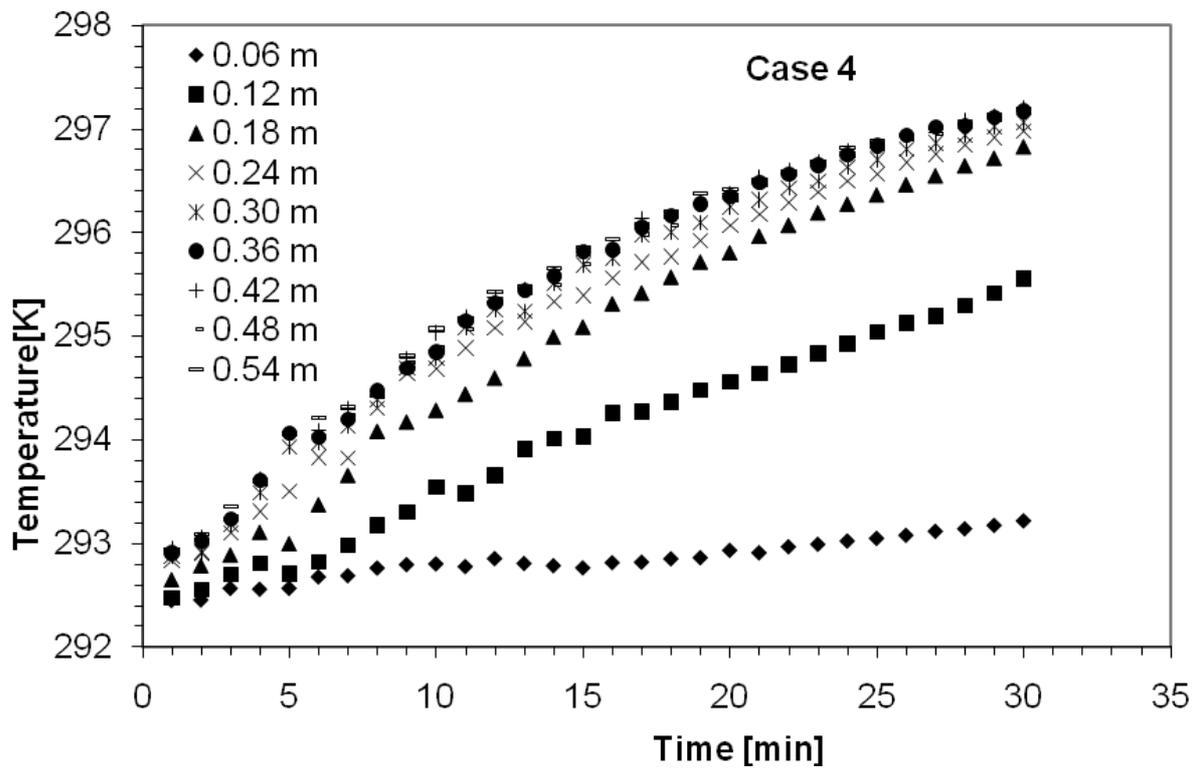

**Fig. 4d**

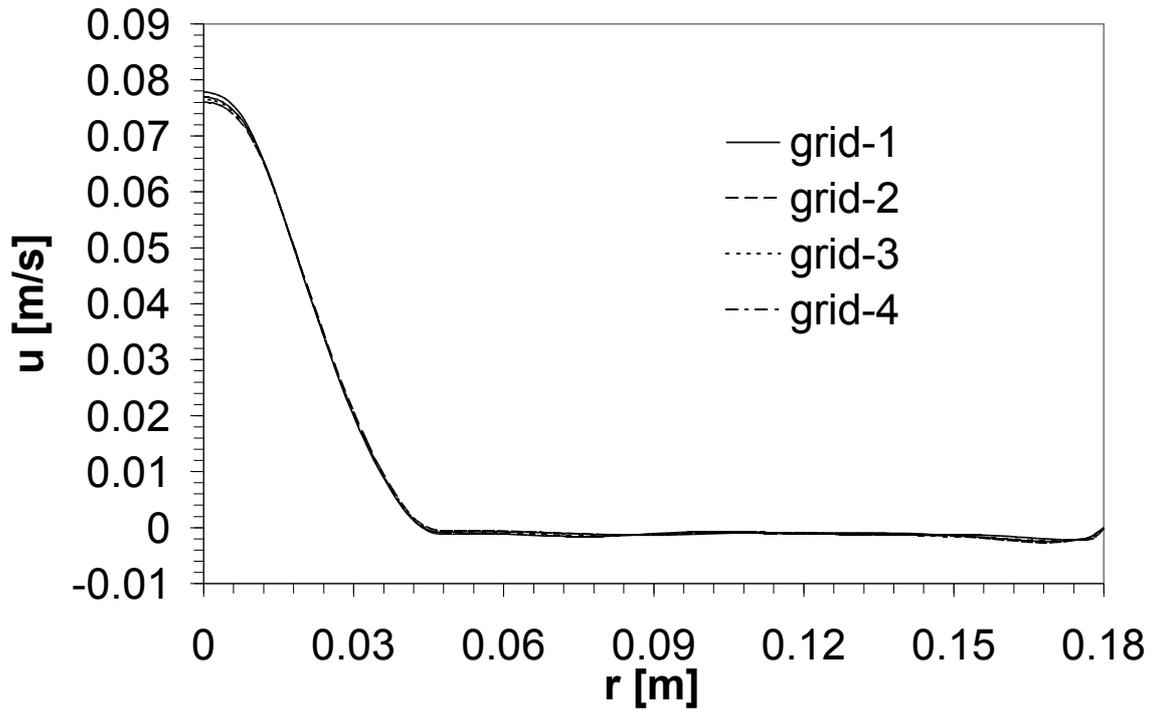

**Fig. 5a**

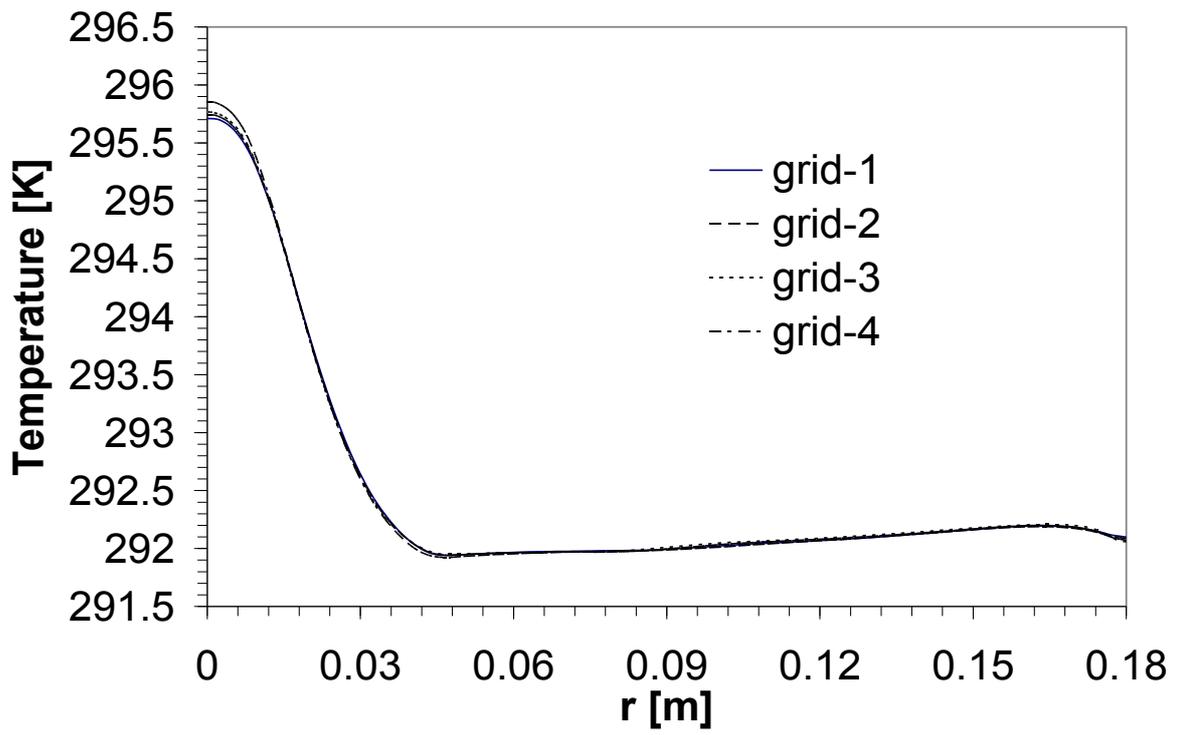

**Fig. 5b**

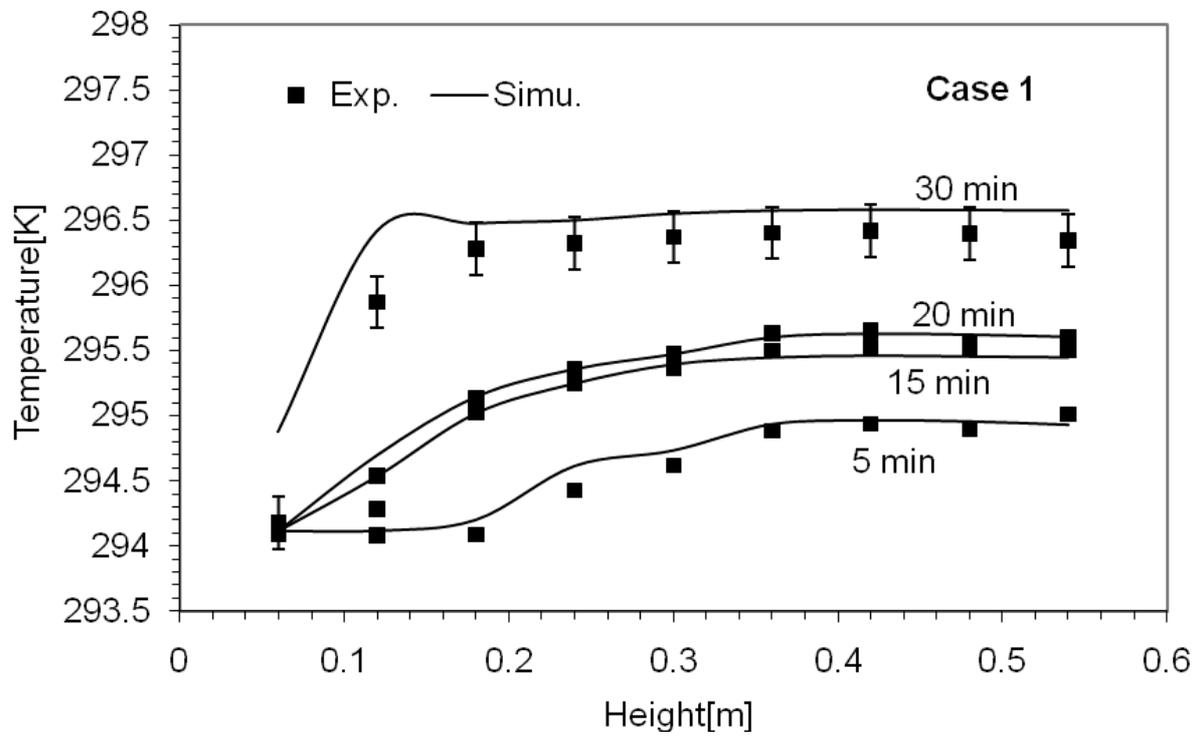

**Fig. 6a**

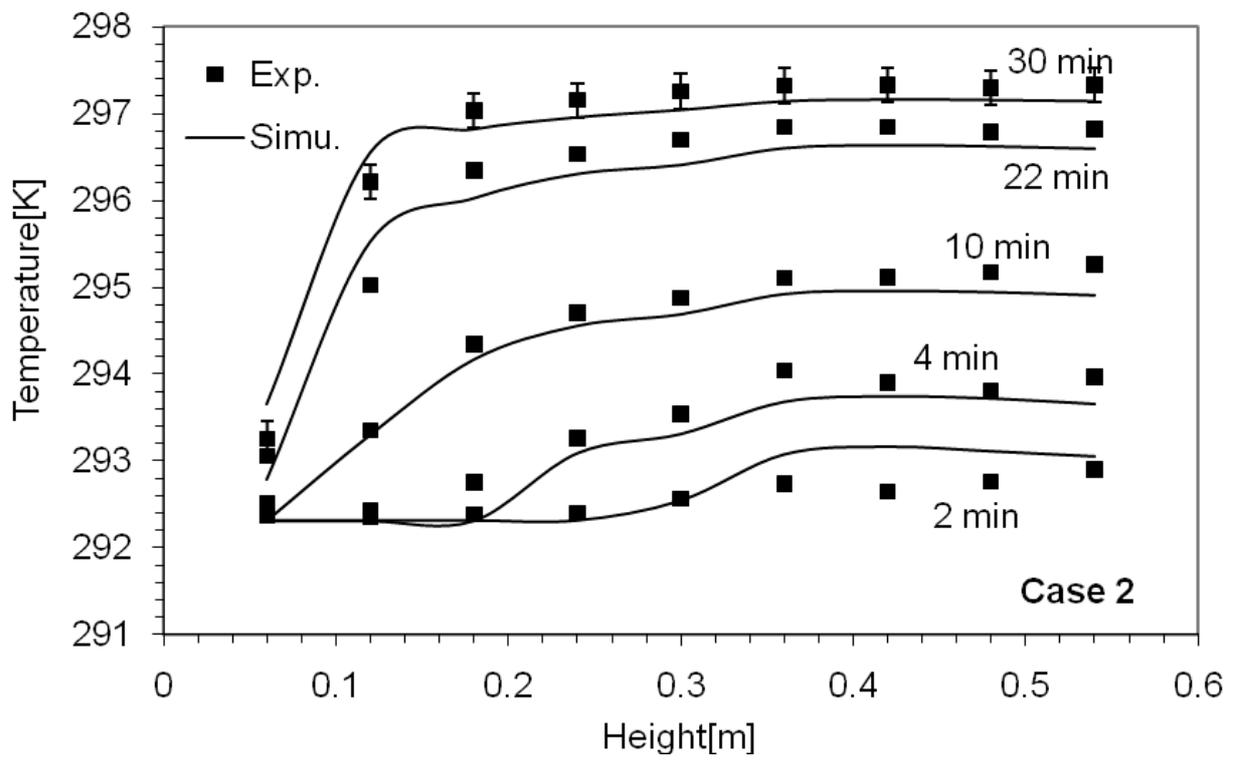

**Fig. 6b**

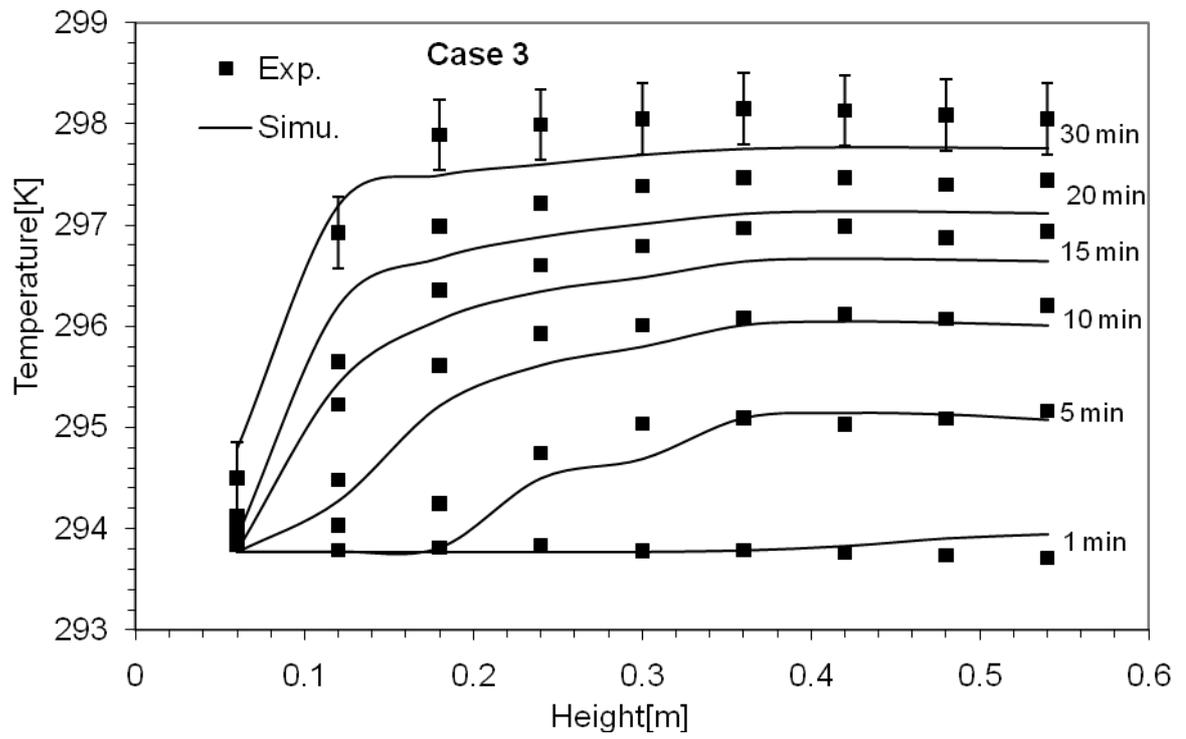

**Fig. 6c**

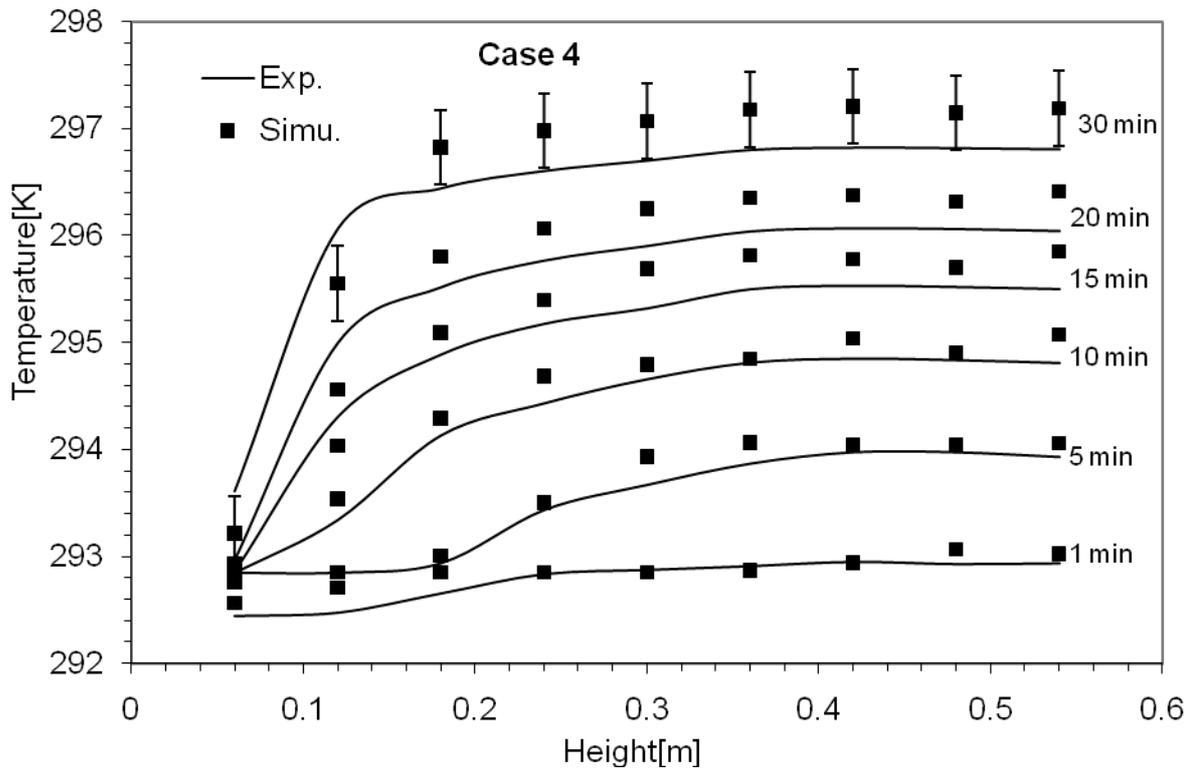

**Fig. 6d**

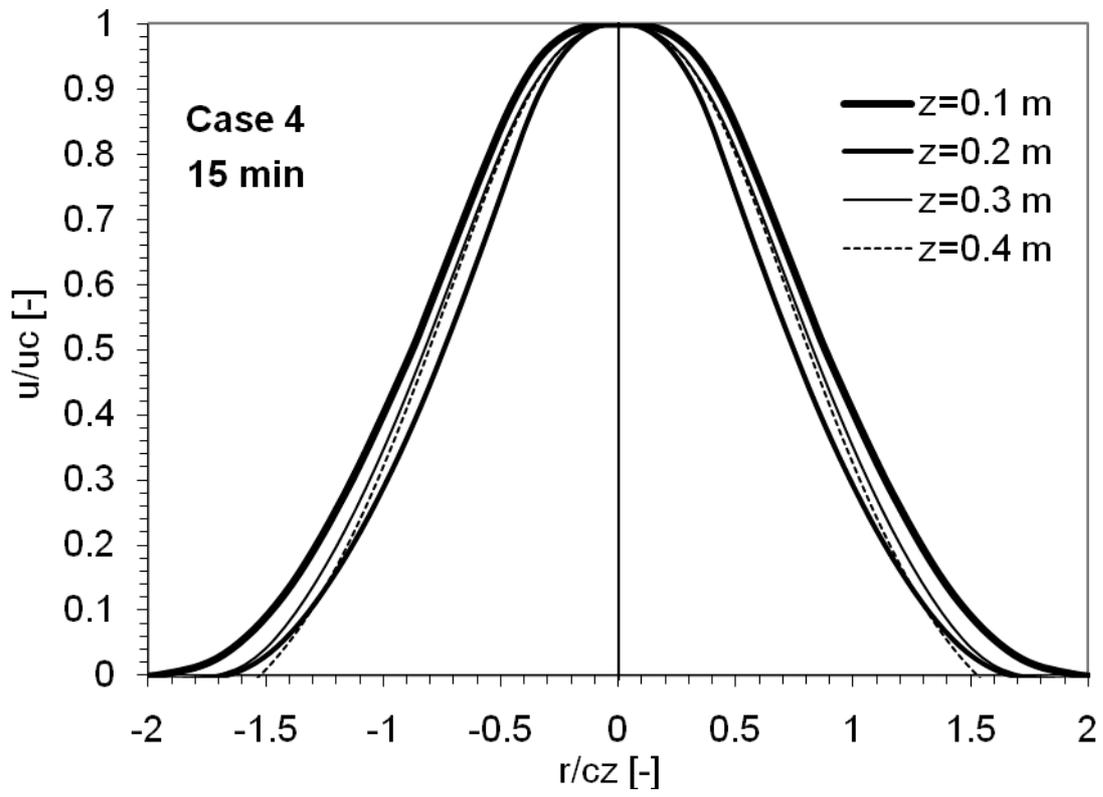

**Fig. 7**

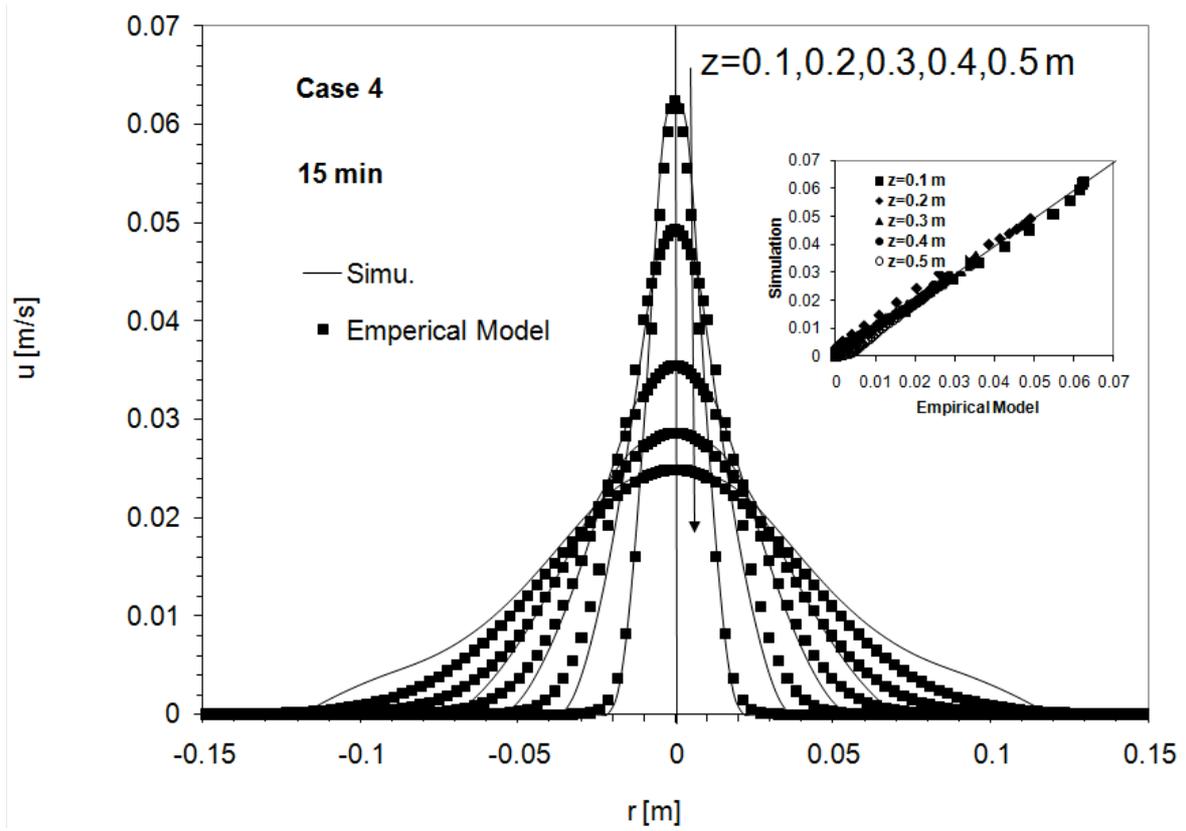

**Fig. 8**

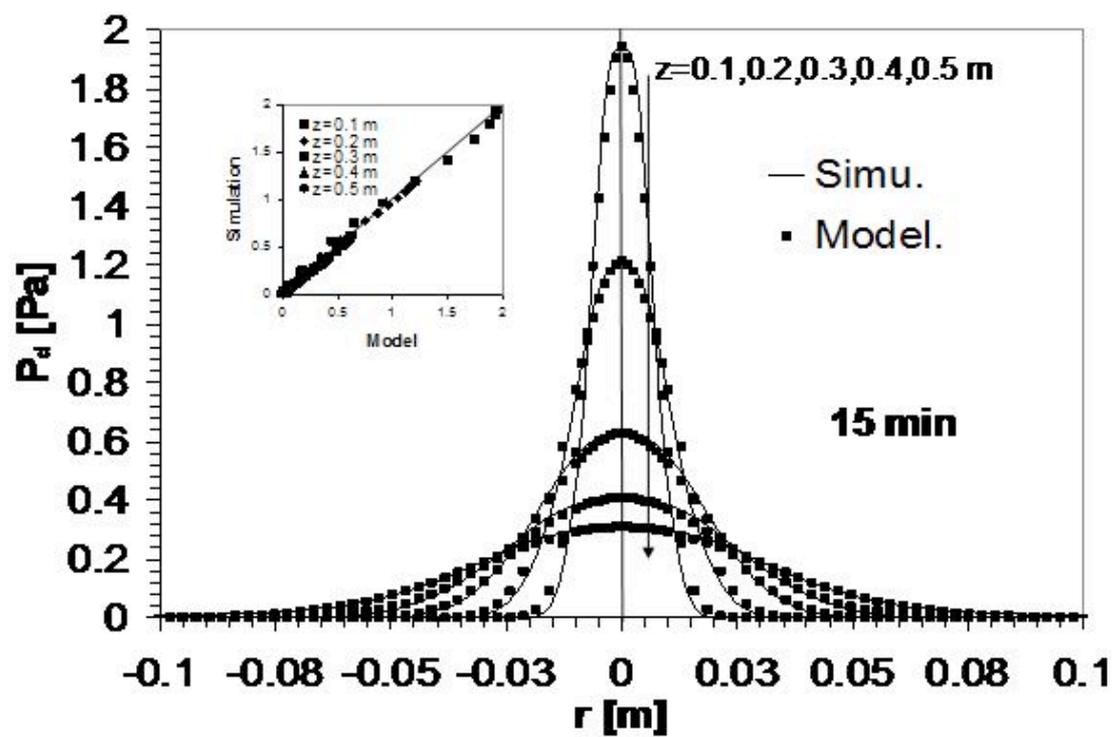

**Fig. 9**

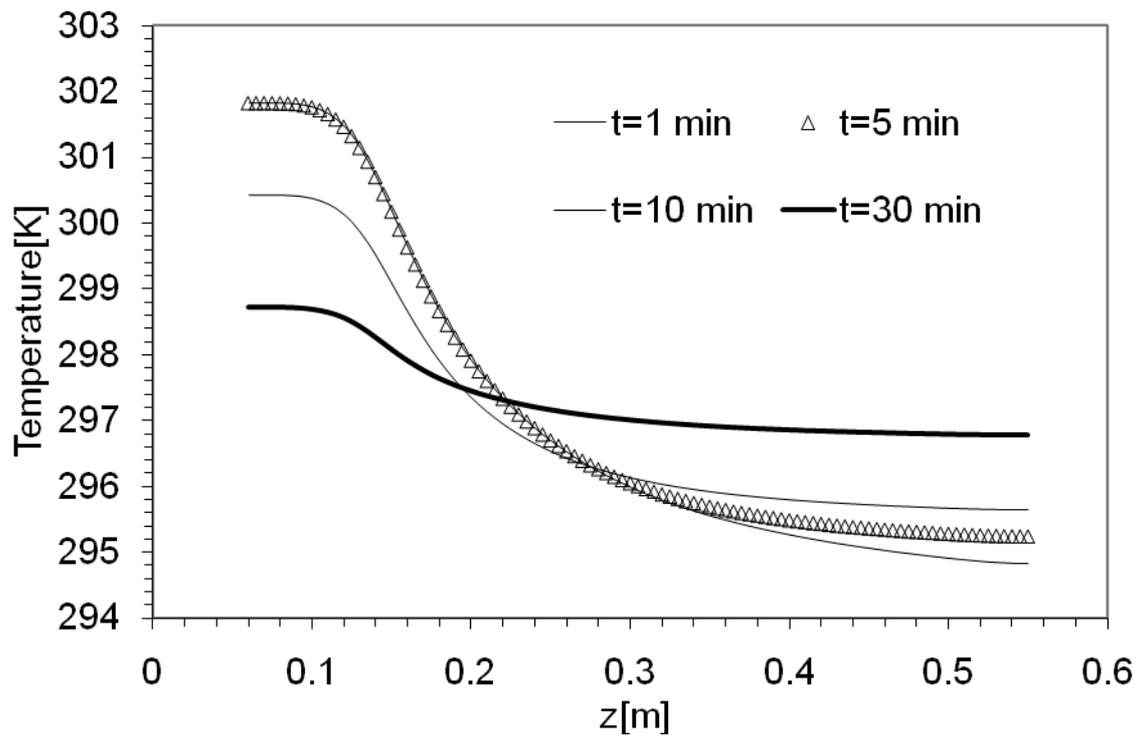

**Fig. 10**

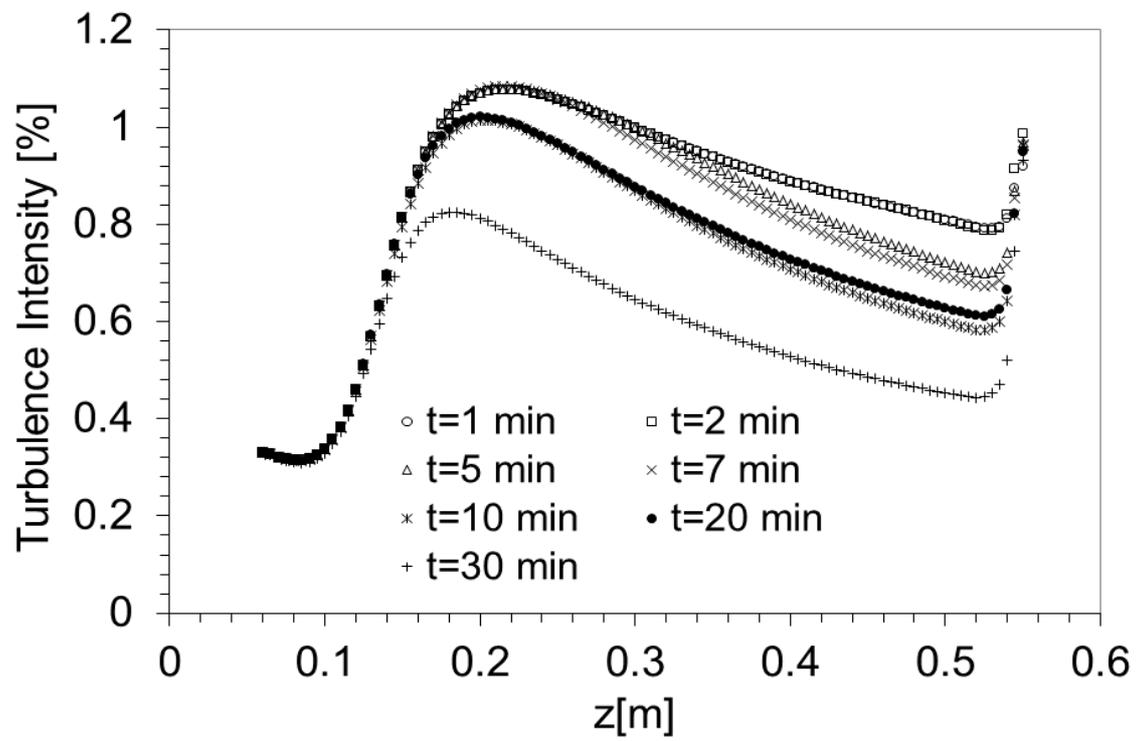

**Fig. 11**

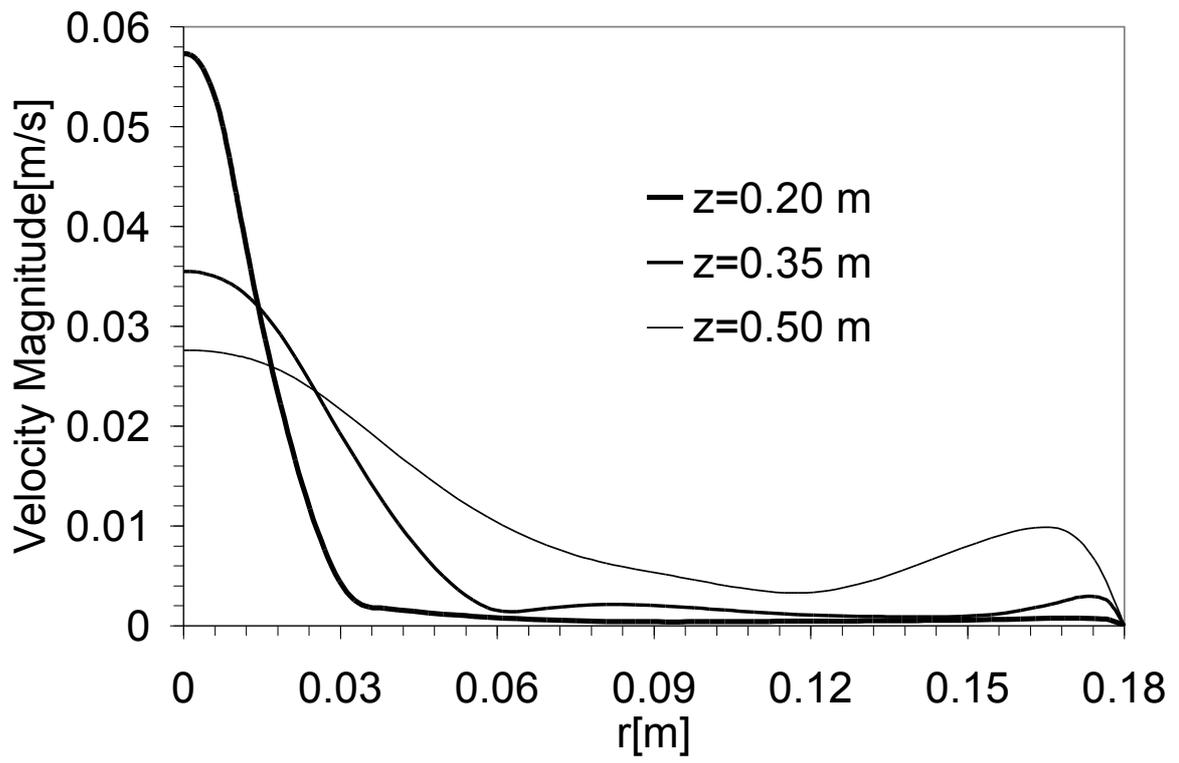

**Fig. 12**

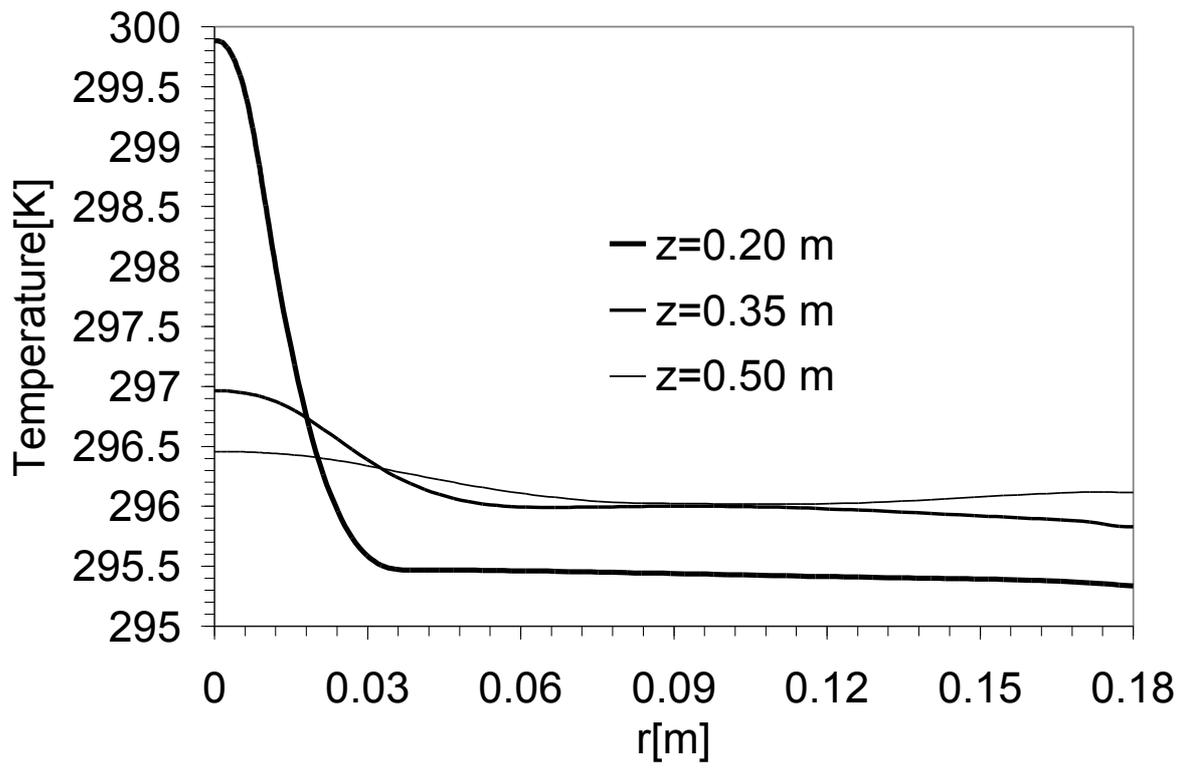

**Fig. 13**

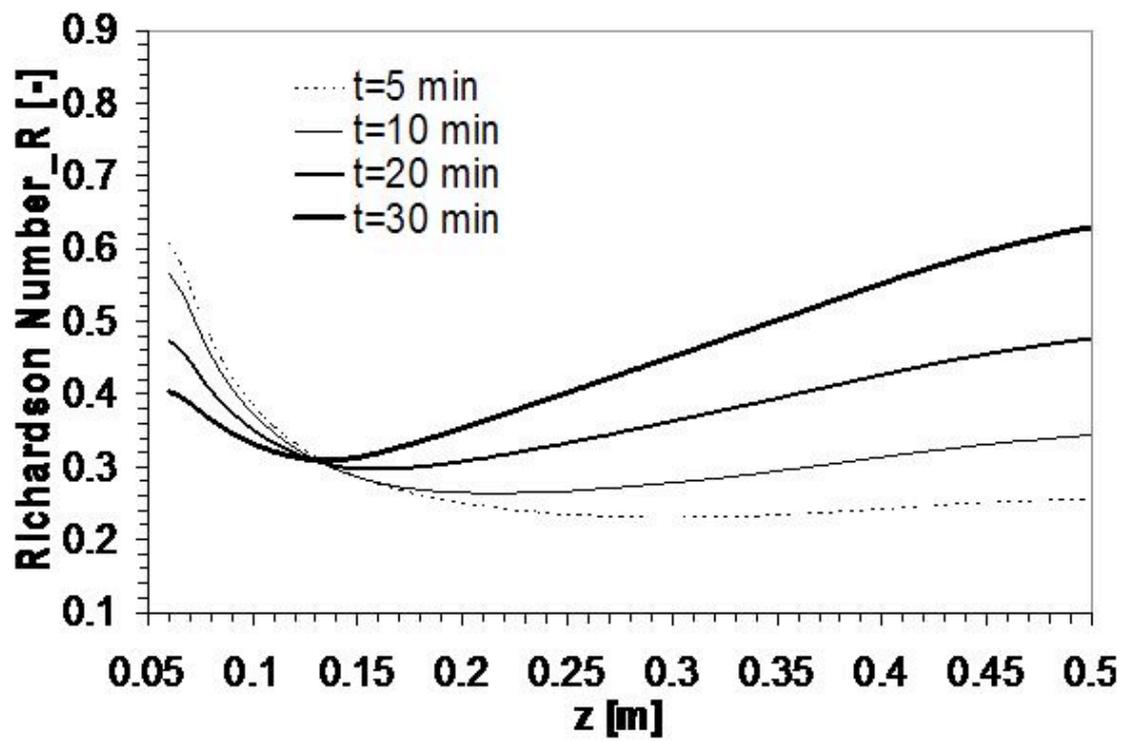

**Fig. 14**

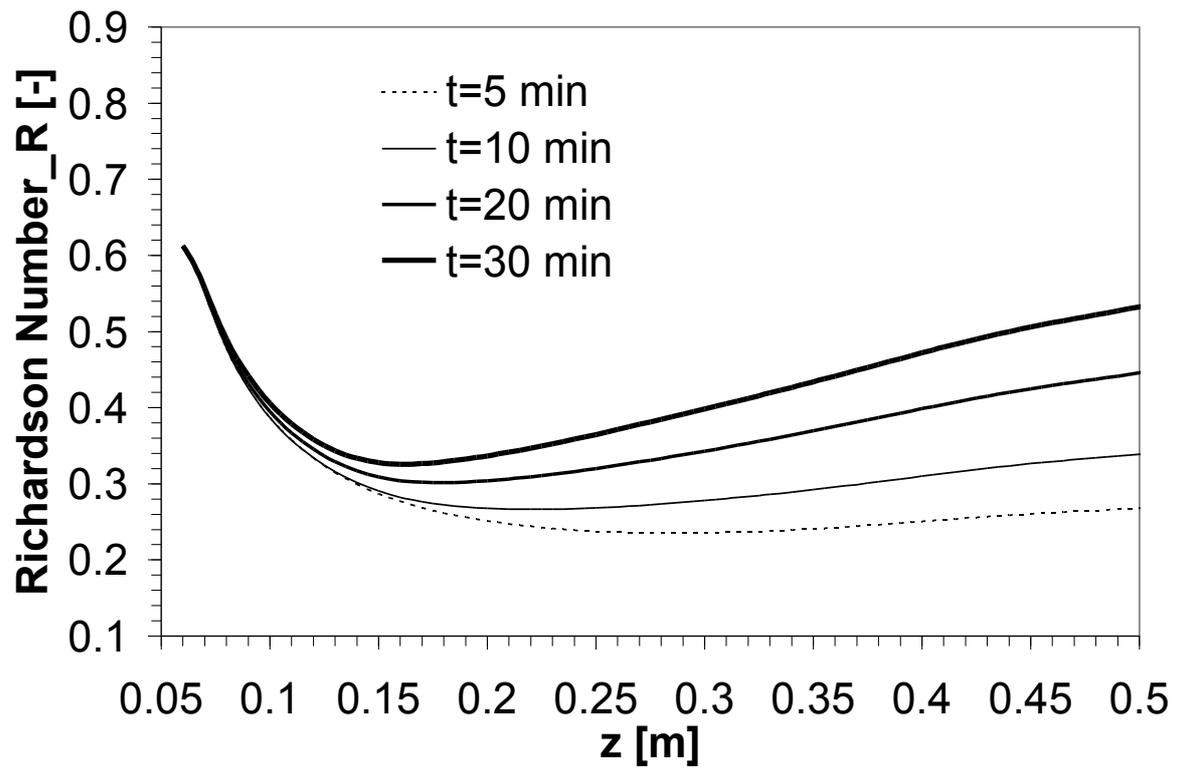

**Fig. 15**

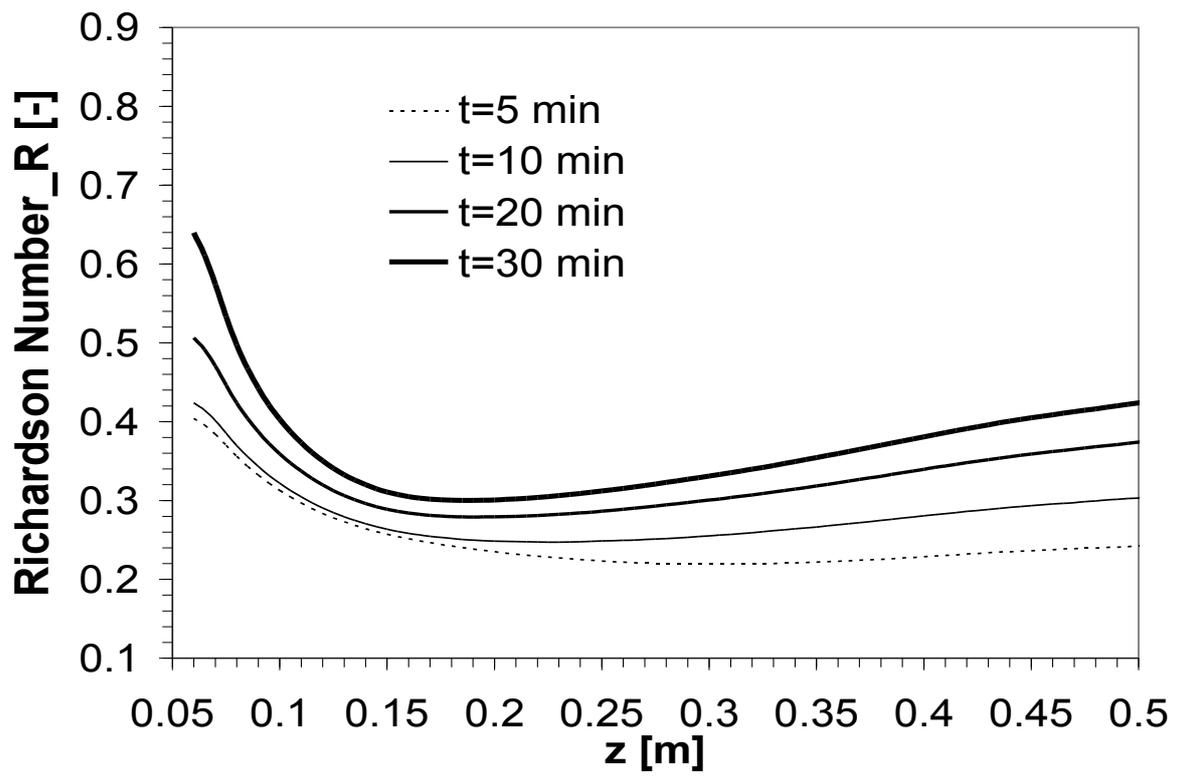

**Fig. 16**

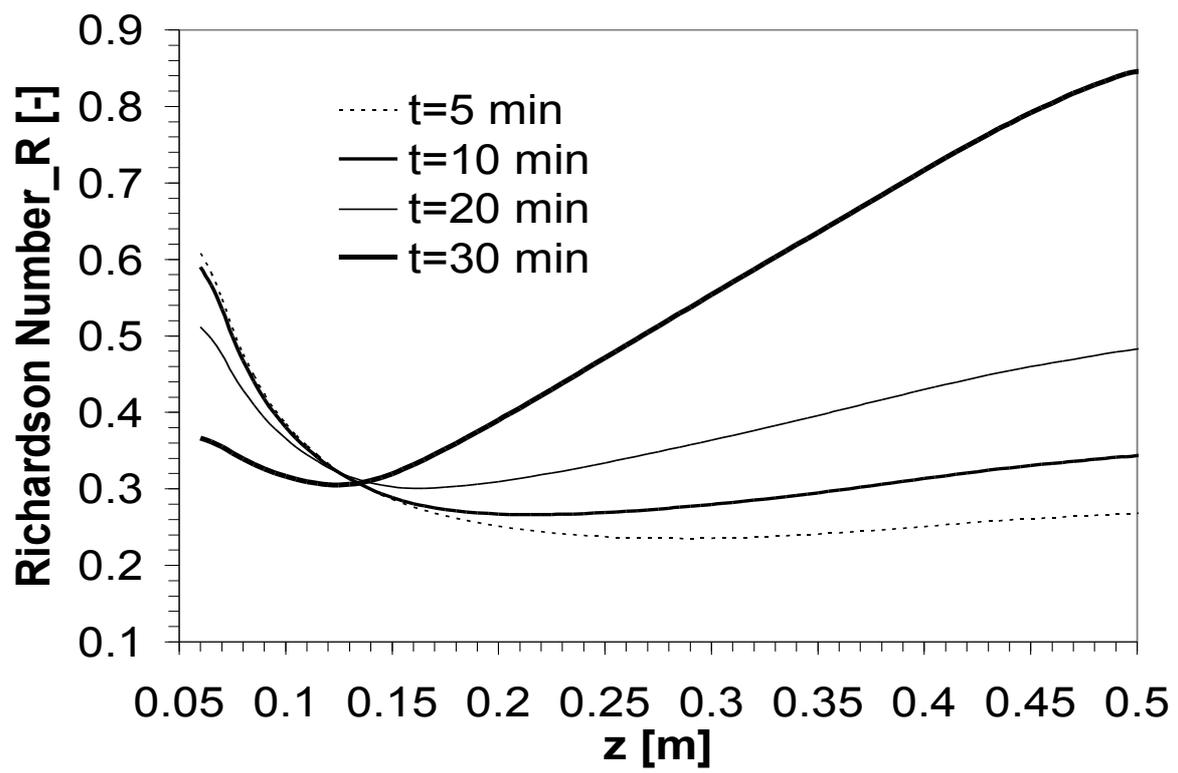

**Fig. 17**